\documentclass[acmsmall,screen,nonacm]{acmart}
\usepackage[utf8]{inputenc}
\usepackage{colortbl}
\usepackage{graphicx}
\usepackage{amsmath}
\usepackage{algpseudocode}
\usepackage{algorithm}

\usepackage{amssymb}
\usepackage{tikz}
\usetikzlibrary{backgrounds,fit}
\usepackage{pgfplots}
\pgfplotsset{compat=newest}
\usepackage{float}
\usepackage{placeins}
\usepackage{pdflscape}
\usetikzlibrary{arrows.meta, positioning, calc, shapes.geometric}
\usepackage{enumitem} 
\usepackage{tabularx}
\usepackage{url}
\usepackage{array,ragged2e,tabularx,booktabs}
\newcolumntype{Y}{>{\RaggedRight\arraybackslash}X} % left-aligned X
\newcolumntype{P}[1]{>{\raggedright\arraybackslash}p{#1}} 
\usepackage{enumitem}  
\setlist[itemize]{leftmargin=*, labelsep=0.5em}
\usepackage{fancyhdr}
\authorsaddresses{}

\AtBeginDocument{%
  }

\setcopyright{none}
\acmJournal{CSUR}
\begin{document}
\fancypagestyle{firstpagestyle}{%
  \fancyhf{}%
  \fancyfoot[C]{\footnotesize Pre-print. Submitted to \emph{ACM Computing Surveys} (under review) --- 1 July 2025.}%
}

\title{Architectural Backdoors in Deep Learning: A Survey of Vulnerabilities, Detection, and Defense}


\author{Victoria Childress}
  \orcid{0000-0001-8941-6675}
\author{Josh Collyer}
  \orcid{0009-0002-0830-5282}
\author{Jodie Knapp}
  \orcid{0000-0002-5929-2015}
\affiliation{%
  \institution{The Alan Turing Institute}
  \city{London}
  \country{UK}
}

\renewcommand{\shortauthors}{Childress \emph{et al.}}

\begin{abstract}
Architectural backdoors pose an under-examined but critical threat to deep neural networks, embedding malicious logic directly into a model’s computational graph. Unlike traditional data poisoning or parameter manipulation, architectural backdoors evade standard mitigation techniques and persist even after clean retraining. This survey systematically consolidates research on architectural backdoors, spanning compiler-level manipulations, tainted AutoML pipelines, and supply-chain vulnerabilities. We assess emerging detection and defense strategies, including static graph inspection, dynamic fuzzing, and partial formal verification, and highlight their limitations against distributed or stealth triggers. Despite recent progress, scalable and practical defenses remain elusive. We conclude by outlining open challenges and proposing directions for strengthening supply-chain security, cryptographic model attestations, and next-generation benchmarks. This survey aims to guide future research toward comprehensive defenses against structural backdoor threats in deep learning systems.
\end{abstract}




\maketitle

    \section{Introduction}
    \label{intro}
    
    Architectural backdoors pose a persistent and under-examined threat to deep neural networks. By embedding malicious logic directly into the computational graph, they evade traditional defenses and persist even after retraining. Because the exploit is hard‑wired (e.g., as an extra branch, gating layer, or rerouted edge), it typically remains dormant and can persist after weight reinitialization. These structures only activate in response to attacker-defined inputs or trigger patterns. By encoding the trigger logic directly in new or rewired operators, architectural backdoors are immune to standard mitigation techniques like data cleansing, weight resets, or fine-tuning alone. Conceptually, such a structural exploit is akin to a hardware Trojan gate hidden in an integrated circuit: the malicious sub-graph lies dormant until a secret trigger activates it~\cite{tehranipoor2010survey, warnecke2023evil}. Fine-grained pruning can disable simple single-path backdoors \cite{Liu2018FinePruning}, but sophisticated or distributed sub-graphs still evade such defenses \cite{min2024gtba}. Prior backdoor surveys have largely focused on data or weights; by contrast, we illuminate architectural backdoors as an emerging threat class requiring dedicated study. Earlier reviews focus on data- or weight-level Trojans \cite{Li2022BackdoorSurvey, Bai2025BackdoorSurvey, Zhao2025LLMBackdoors}. Instead, we will synthesize the emerging literature on architectural backdoors and their unique detection and mitigation challenges.
\noindent
    Recent work has revealed that architectural backdoors can do more than trigger misclassifications: by exploiting within-batch inference they can leak or manipulate the outputs of other users who share the same batch, breaking the isolation guarantees of large-scale model serving \cite{Kuchler2025BatchSteal}. This privacy-critical scenario extends the threat surface first highlighted by Bober-Irizar et~al.~\cite{Bober-Irizar2023}, who coined the term Model Architectural Backdoor (MAB) for malicious logic wired directly into a network’s computational graph.
    
    \begin{table}[htbp]
      \caption{\textbf{Common backdoor types}}
      \label{tab:backdoor-types}
      \centering
      \footnotesize
      % one fixed-width column + one flexible column
      \begin{tabularx}{\linewidth}{P{3.0cm} Y}
        \toprule
        \textbf{Type} & \textbf{Description} \\
        \midrule
        \textbf{Data poisoning} &
          Injecting triggers into training samples (often mislabeled) to
          activate hidden misbehavior at inference \\ \midrule
        \textbf{Weight backdoor} &
          Manipulating model parameters or fine-tuning with malicious objectives
          to cause trigger-based misclassifications \\ \midrule
        \textbf{Architectural} &
          Modifying the network’s structure (e.g., extra subgraphs, backdoor
          layers) that persist even under clean retraining \\
        \bottomrule
      \end{tabularx}
    \end{table}
\medskip   
\noindent  Structural backdoors resist standard removal: once a malicious sub-graph is embedded in the architecture, it survives weight re-initialization and clean retraining, leaving safety-critical models exposed. Automated machine learning (AutoML) and neural architecture search (NAS) pipelines hide such logic inside increasingly complex topologies, making detection harder.  Real-world evidence now exists: \emph{AI}’s \emph{Guardian} scanner has examined 4.47 million model versions in 1.41 million Hugging Face repositories and has flagged 352,000 unsafe or suspicious issues across 51,700 distinct models, including the popular \emph{SoccerTwos} ONNX file, for architectural-backdoor–like sub-graphs \cite{ProtectAI2025SixMonthReport}. Recent work such as HiddenLayer’s \emph{Shadow Logic} likewise shows trigger-activated misbehavior persisting even after full ImageNet retraining \cite{hiddenlayer2024}. In a safety-critical setting, embedded triggers might remain dormant until activated by carefully crafted stimuli, leading to catastrophic failures at the worst possible moments. The combination of stealth and persistence thus demands dedicated study of architectural backdoors, filling a gap left by previous surveys on data- and weight-based backdoors.
    
    \begin{figure}[ht]
    \centering
    \begin{tikzpicture}[%
      font=\footnotesize,
      >={Stealth},
      shorten >=1pt,
      node distance=0.6cm and 1.0cm,
      align=center,
      attack/.style={draw=red, rectangle, minimum width=3.3cm,
                     minimum height=0.7cm, fill=red!15},
      clean/.style={draw, rectangle, minimum width=3.3cm,
                    minimum height=0.7cm},
      normalarrow/.style={->},
      attackarrow/.style={->, red}
    ]
    
    % ---------- (a) Data-poisoning ----------------
    \node at (0.0, 1.0) {\textbf{(a) Data Poisoning}};
    \node[attack] (poisoneddata) {Poisoned Data};
    \node[clean, below= of poisoneddata] (cleanarch) {Clean Architecture};
    \node[clean, below= of cleanarch] (normalweights) {Normal Weights};
    
    \draw[normalarrow] (poisoneddata.south) --
          (cleanarch.north) node[midway,right,xshift=2pt] {\small training};
    \draw[normalarrow] (cleanarch.south) --
          (normalweights.north) node[midway,right,xshift=2pt] {\small std.\ init};
    
    % ---------- (b) Weight backdoor ----------------
    \node at (4.75, 1.0) {\textbf{(b) Weight Backdoor}};
    \node[clean, xshift=4.75cm] (cleandata) {Clean Data};
    \node[clean, below= of cleandata] (cleanarch2) {Clean Architecture};
    \node[attack, below= of cleanarch2] (hackedweights) {Hacked Weights};
    
    \draw[normalarrow] (cleandata.south) --
          (cleanarch2.north) node[midway,right,xshift=2pt] {\small training};
    \draw[attackarrow] (cleanarch2.south) --
          (hackedweights.north) node[midway,right,xshift=4pt] {\small attacker alters};
    
    % ---------- (c) Architectural backdoor ---------
    \node at (9.5, 1.0) {\textbf{(c) Architectural Backdoor}};
    \node[clean, xshift=9.5cm] (cleandata2) {Clean Data};
    \node[attack, below= of cleandata2] (malarch) {Malicious Architecture};
    \node[clean, below= of malarch] (normalweights2) {Normal Weights};
    
    \draw[normalarrow] (cleandata2.south) --
          (malarch.north) node[midway,right,xshift=2pt] {\small training};
    \draw[normalarrow] (malarch.south) --
          (normalweights2.north) node[midway,right,xshift=2pt] {\small std.\ init};
    \end{tikzpicture}
    
    \caption{\label{fig:backdoor-compare}Three main backdoor insertion points.  
    (a)~data poisoning, (b)~weight backdoor, and (c)~architectural manipulation.  
    Attacker-altered elements are shown in red.}
    \Description{Three side-by-side mini diagrams compare (a) data poisoning, (b) weight backdoors, and (c) architectural backdoors. Red-tinted boxes and red arrows highlight where the adversary intervenes in each scenario.}
    \end{figure}
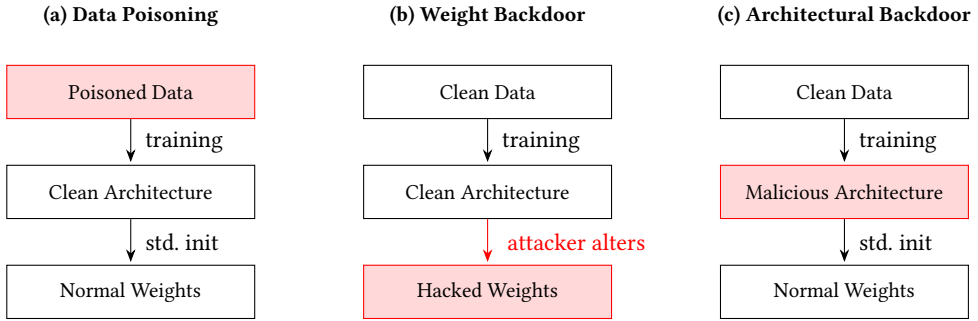

\subsection{Contributions and Organization}
\label{sec:intro_contributions}

Our main contributions are:

\begin{itemize}
    \item \textbf{Comprehensive Taxonomy of Architectural Backdoors.} We categorize different architectural attacks, including sub-network modifications, compiler-level backdoor insertion, and AutoML vulnerabilities, detailing their distinct threat models and stealth properties.

    \item \textbf{Detection and Mitigation Strategies.} We survey existing techniques (e.g., static graph inspection, dynamic fuzzing, formal verification, subgraph pruning) and assess their efficacy against architecture-centric backdoors.

    \item \textbf{Open Challenges and Future Directions.} We highlight research gaps such as supply-chain security, robust model verification at scale, and multi-branch triggers, emphasizing the need for comprehensive defenses that go beyond data or weight checks.
\end{itemize}

\noindent
\textbf{§\ref{sec:background}} reviews classical backdoor mechanisms and clarifies how architectural attacks fit within the broader landscape. \textbf{§\ref{sec:taxonomy}} presents our taxonomy of architectural backdoors. \textbf{§\ref{sec:detection}} delves into detection techniques, followed by mitigation and model repair strategies in \textbf{§\ref{sec:mitigation}}. \textbf{§\ref{sec:benchmarks}} surveys available benchmarks, datasets, and empirical evaluations for assessing architectural backdoors, and \textbf{§\ref{sec:future}} discusses open challenges and future directions. Finally, \textbf{§\ref{sec:conclusion}} concludes with implications for securing AI models against architectural backdoors.

\section{Background and Related Work}
\label{sec:background}

A backdoor can be implanted (i) in the training data, (ii) in the model weights, or (iii) in the network architecture. The remainder of this section explains why architectural backdoors demand separate treatment and motivates the taxonomy that follows.

\subsection{Classical vs.\ Architectural Backdoors}
\label{sec:classical-arch}

\noindent
\noindent\paragraph{Classical (Data- and Weight-Based) backdoors}
Most early backdoor research, as surveyed in prior reviews~\cite{Bai2025BackdoorSurvey,Li2022BackdoorSurvey,Zhao2025LLMBackdoors,Li2020Survey}, centered on manipulating either training data or model weights.  
A prototypical data-poisoning approach is \emph{BadNets}~\cite{Gu2017badnets}, which injects small patterns into a subset of training samples, causing misclassification only when the trigger is present. Meanwhile, weight-based attacks~\cite{Liu2018trojan} alter parameters (e.g., via fine-tuning or partial re-initialization) to produce attacker-chosen outputs upon detecting a secret trigger \cite{li2021backdoor, li-etal-2021-backdoor}. These methods may have minimal footprints in weights or data, making them difficult to prune. Recent work by Goldwasser \emph{et al.}~\cite{Goldwasser2022Undetectable} demonstrated the existence of undetectable weight-edited backdoors, where adversarial parameter edits are computationally indistinguishable from clean weights. These attacks blur the line between whitebox and blackbox threat models, as even full model access may not enable reliable detection.
Classical backdoor defenses often focus on spotting anomalous activations or triggers (e.g., neuron pruning, trigger inversion), which can be partially effective if the backdoor is embedded in data or parameters. However, such methods largely assume suspect training samples or suspicious weights can be identified and removed; that assumption fails if the malicious logic is baked into the graph itself.

\begin{figure}[ht]
\centering
\begin{tikzpicture}[%
  font=\footnotesize,
  >={Stealth},
  shorten >=1pt,
  node distance=1.2cm and 1.6cm,
  align=center,
  normalbox/.style={draw, rectangle, minimum width=2.0cm,
                    minimum height=0.35cm, fill=gray!15},
  normalarrow/.style={->, draw=gray!70},
  hiddenbox/.style={draw=red, rectangle, thick,
                    minimum width=2.3cm, minimum height=0.9cm,
                    fill=red!10},
  hiddenarrow/.style={->, red, dashed, thick}
]

% ---- ordinary forward path (grey) ----
\node[normalbox, minimum width=1.6cm] (input) {Input};
\node[normalbox, right= of input, xshift=0.6cm] (convstack) {Conv / Pool};
\node[normalbox, right= of convstack, xshift=0.7cm] (fc) {Final Class};

\draw[normalarrow] (input.east) -- (convstack.west)
      node[midway, above] {\scriptsize normal path};
\draw[normalarrow] (convstack.east) -- (fc.west);

% ---- hidden trigger branch (red) ----
\node[hiddenbox, below=0.45cm of convstack] (triggerdet)
      {\textcolor{red}{Trigger Detector}\\\scriptsize(e.g.\ checkerboard)};

\draw[hiddenarrow] (input.south) |- (triggerdet.west)
      node[pos=0.38, right, xshift=3pt] {\scriptsize skip};
\draw[hiddenarrow] (triggerdet.east) -| (fc.south)
      node[midway, left, xshift=-2pt, yshift=4pt] {\scriptsize override};

\node[below=0.2cm of triggerdet, text width=13cm, align=center, red]
      {\footnotesize Activates only when the trigger pattern is present; otherwise produces zero.};

\end{tikzpicture}

\caption{\label{fig:triggesschematic}\textbf{Hidden-branch backdoor schematic.} (adapted from Bober-Irizar \emph{et al.} \cite{Bober-Irizar2023}) A light-gray normal path processes inputs conventionally, while a red hidden branch bypasses it.  
When the trigger pattern is detected, the branch overrides the main features, forcing the attacker-chosen output.}
\Description{Block diagram of a neural network. The normal forward path is shown in light gray, whereas a bold red hidden branch bypasses the normal layers and activates only on a special trigger pattern to override the final classifier.}
\end{figure}
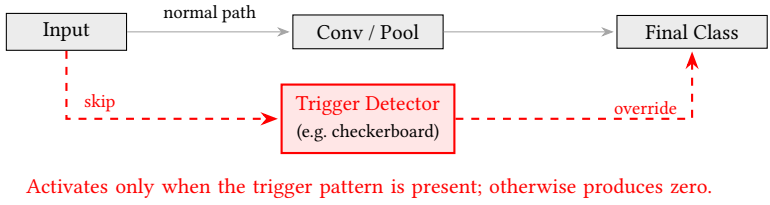

\paragraph{Architectural backdoors}
Recent works~\cite{Bober-Irizar2023,Clifford2024} show that structural modifications survive weight re-initialization, retraining on clean data, or standard pruning strategies. A controlled study by Langford \emph{et~al.}~\cite{Langford2025} demonstrates this empirically: even after full weight re-initialization
followed by clean \emph{ImageNet} retraining, their architectural Trojan preserved a $96.2\%$ attack-success rate, whereas a baseline \emph{BadNets} weight backdoor fell below $2\%$. Earlier proof-of-concept work by Qi \emph{et al.} demonstrated that a tiny malicious sub-network can even be inserted after training, at deployment time, yet still pass unnoticed and retain 100 \% attack success once triggered~\cite{Qi2023SRA}. This persistence arises because the malicious branch is weight-agnostic: its routing or gating logic is hard-wired into the graph topology rather than stored in learned parameters.  Re-initializing or retraining the weights leaves those fixed routes untouched, so the backdoor re-emerges as soon as training converges. By wiring malicious routing or gating logic directly into the computational graph, attackers ensure it can be selectively activated by triggers.  Because this backdoor is not encoded in the parameters or the data, standard defenses (data cleansing, weight resets) cannot remove the malicious subgraph.  Figure~\ref{fig:triggesschematic} offers a schematic example of a hidden branch that remains dormant until a certain pattern is found, then overrides the final classifier.  Similar structures have now been demonstrated in large-language models~\cite{Zhao2025LLMBackdoors,Miah2024ArchLLM,xiang2024badchain}, confirming a cross-modal threat.  Backdoor attacks now extend beyond image classifiers to diffusion models~\cite{Chou2023BadDiffusion} and LLM-powered recommender systems~\cite{Ning2025PScanner}.

\newpage
\subsection{Motivation for Architectural Threats}
\label{sec:motivation-struct}

Architectural backdoors have gained attention for several reasons:

\begin{itemize}[leftmargin=1.4em,itemsep=2pt]
    \item \textbf{Complex Model Design.} As models grow larger (e.g., vision transformers, large language models (LLMs)), subtle architectural manipulations become harder to detect, providing cover for hidden subgraphs \cite{pang2022,pang2023dark}.
    \item \textbf{Supply-Chain Exposure.} A compromised compiler, converter, or accelerator can inject logic \emph{after} training, as foreshadowed by Ken~Thompson’s classic ``Reflections on Trusting~Trust’’ compiler Trojan\,\cite{thompson1984reflections} and demonstrated in modern ML contexts by \emph{Shadow Logic} and \emph{ImpNet}\,\cite{hiddenlayer2024,Clifford2024}. \emph{ImpNet}, in particular, exemplifies a compiler-level backdoor that is both imperceptible and black-box undetectable: by inserting steganographic trigger detectors at the graph Intermediate Representation (IR) level, the attacker creates a dormant conditional pathway that activates only on extremely specific binary input patterns. The model’s accuracy on clean inputs remains unchanged, and the trigger cannot be synthesized or detected without reverse-engineering the compiled machine code \cite{Clifford2024}. Related hardware-Trojan risks are surveyed in \cite{tehranipoor2010survey}.

    \item \textbf{Stealth and Persistence.} Because these backdoors reside in the model topology, many classical defenses (e.g.\ weight pruning or trigger inversion) cannot remove them.
    
    \item \textbf{AutoML Bias.}  NAS often yields shallower, wider architectures with extensive skip-connections \cite{Pang2022AutoMLRisk}. Skip connections themselves are a perfectly benign and widely used design feature (e.g. \emph{ResNets}); they become a security concern only when an attacker leverages that extra routing flexibility to embed hidden logic or alternative pathways.  Empirical analyses by Pang \emph{et~al.}, \cite{Pang2022AutoMLRisk,pang2023dark} show that such automatically searched models are often more vulnerable in empirical benchmarks to adversarial, poisoning, and Trojan attacks than carefully engineered CNNs. This “search-bias” gives attackers a foothold that is largely absent in hand-crafted architectures.
    
    \item \textbf{Real-World Incidents.}  Recent supply-chain breaches, including a dependency hijack in PyTorch nightly builds (Dec 2022), malware-laced models on Hugging Face (Feb 2024), and structural backdoors uncovered in the NIST/IARPA \emph{TrojAI} corpus (Round 15, 2024), a round that, for the first time, pivots the benchmark from input-patch triggers to graph-level structural Trojans, underscoring the field’s growing concern\cite{trojai-round15}, show how easily malicious logic reaches production pipelines. In its first six months, Protect AI’s Guardian scanner examined \textasciitilde4.47 million model versions across 1.41 million repositories and flagged about 352,000 suspicious issues affecting 51,700 models \cite{ProtectAI2025SixMonthReport}.

\end{itemize}

\noindent
Converging trends such as complex graphs, AutoML shortcuts, and porous supply chains set the stage for the structural backdoor attacks surveyed in the following sections. Understanding them is essential to designing benchmarks and defenses that remain robust as models and toolchains evolve. Researchers have also highlighted potential hardware Trojans in machine-learning (ML) accelerators~\cite{Sengupta2025HardwareTrojanML}, which remain invisible to software-level analysis.
\noindent
Deep-learning inference increasingly runs on dedicated chips such as Field-Programmable Gate Arrays (FPGAs), reconfigurable logic fabrics that can be retargeted after manufacture, and Application-Specific Integrated Circuits (ASICs), custom-fabricated devices optimized for a single workload (e.g., Google’s TPU dies). Because these accelerators bypass a general-purpose CPU’s security stack, even a single malicious logic cell in the bitstream can silently re-enable an otherwise neutralized backdoor.
\noindent
Transformers and other sequence models have likewise been shown to be susceptible to hidden gating sub-layers that activate on specific token patterns~\cite{Zhao2025LLMBackdoors,Miah2024ArchLLM}, broadening the scope of structural threats.

\subsection{Key Terminology and Scope}
\label{sec:terminology-scope}

In this survey, we use:

\begin{itemize}
    \item \textbf{Trigger:} A specific input pattern or computational condition that induces malicious outputs. Triggers are designed to be inconspicuous.
    
    \item \textbf{Malicious Subgraph.}  Structural additions or modifications in the computational graph that remain dormant under normal conditions but activate upon a trigger, overriding normal inference \cite{Bober-Irizar2023}. Although many such subgraphs are designed to have no measurable effect until the trigger fires, sophisticated variants can still leak a small bias into the output distribution even in the nominal (untriggered) state, which complicates detection. A concrete early example is \emph{TrojanNet}, which hides an entire classifier inside a host network and triggers it with a tiny pixel key \cite{Guo2021TrojanNet}.

    \item \textbf{Intermediate Representation (IR)}: A hardware- or framework-specific, low-level graph of operations that sits between a high-level model definition (e.g., PyTorch or TensorFlow code) and the final machine-code or micro-kernel instructions, allowing compilers to optimize and schedule the network for a particular accelerator. ONNX (Open Neural Network Exchange) is an open, hardware-neutral IR standard that captures a model’s computational graph, including operators, shapes and data types, so that models trained in one framework can be ported, optimized and executed across many runtimes and devices without re-implementation.
    \item \textbf{Compiler-Level Backdoor:} Logic inserted at the compilation or model-export stage (e.g., into ONNX or other IR), hidden from source-level review~\cite{Clifford2024}. 
    \item \textbf{NAS/AutoML Backdoors:} Malicious architectures generated by tampering with automated architecture search processes, persisting even when trained on clean data~\cite{pang2022,pang2023dark}.
\end{itemize}

\subsection{Major Architectural Attack Mechanisms}
\label{sec:architectural-attacks}

Architectural backdoors can be implanted at various stages, such as model design, compilation, or AutoML-based generation. §\ref{sec:taxonomy} systematically categorizes these mechanisms, including their threat models and detection challenges.

\begin{figure}[ht]
    \centering
    
\begin{tikzpicture}[%
   font=\footnotesize,
   >={Stealth},
   node distance=1.5cm and 1.5cm,
   align=center,
   boxy/.style={rectangle, draw, minimum width=2.4cm, minimum height=0.8cm, rounded corners}
]

% Nodes
\node[boxy] (design) {Model Design \\(e.g. Arch Code)};
\node[boxy, right=of design] (autoML) {AutoML/\\NAS Pipeline};
\node[boxy, right=of autoML] (compile) {Compilation/\\Graph Opt.};
\node[boxy, right=of compile] (deploy) {Deployment};

% Arrows
\draw[->] (design.east) -- (autoML.west);
\draw[->] (autoML.east) -- (compile.west);
\draw[->] (compile.east) -- (deploy.west);

% Attack points
\node[draw, red, circle, inner sep=1.5pt, below=0.1cm of design] (attack1) {A};
\node[draw, red, circle, inner sep=1.5pt, below=0.1cm of autoML] (attack2) {B};
\node[draw, red, circle, inner sep=1.5pt, below=0.1cm of compile] (attack3) {C};

\draw[red, dashed, -] (attack1.north) -- (design.south);
\draw[red, dashed, -] (attack2.north) -- (autoML.south);
\draw[red, dashed, -] (attack3.north) -- (compile.south);

\node[below=0.1cm of attack1, text width=2.5cm, align=center] {\footnotesize Malicious \\ open-source code};
\node[below=0.1cm of attack2, text width=2.5cm, align=center] {\footnotesize Compromised \\ AutoML services};
\node[below=0.1cm of attack3, text width=2.5cm, align=center] {\footnotesize backdoored \\ compiler};
% Attack point D - Deployment
\node[draw, red, circle, inner sep=1.5pt, below=0.1cm of deploy] (attack4) {D};
\draw[red, dashed, -] (attack4.north) -- (deploy.south);
\node[below=0.1cm of attack4, text width=2.5cm, align=center] {\footnotesize Tampered \\deployment\\container or weights};

\end{tikzpicture}
 \caption{\label{fig:supply-chain}\textbf{Common attack points in the ML supply chain.} }
\Description{Horizontal supply-chain timeline, source code, AutoML/NAS, compiler/export, deployment, annotated with red lightning icons to highlight where architectural backdoors can be inserted.}
\end{figure}
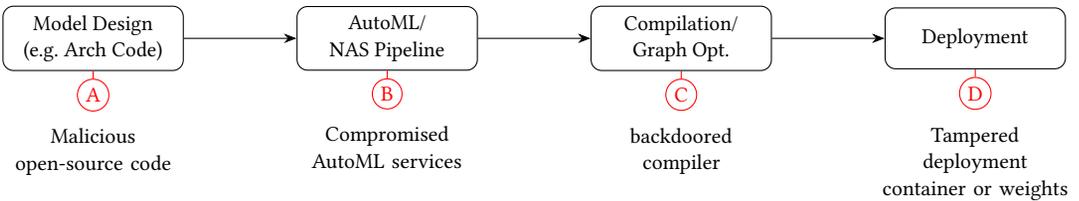

\subsection{Preliminary Defenses and Outstanding Hurdles}
\label{sec:prelim-defenses}

Basic static graph inspection or dynamic testing may detect obvious backdoor branches~\cite{Bober-Irizar2023,Wang2019}, yet more sophisticated designs (e.g., multi-branch or compiler-level) often evade these checks. Weight pruning and trigger inversion, which work against classical backdoors, rarely address structural logic. Formal verification shows promise but faces scalability and complexity barriers, particularly when verifying compiled architectures that diverge from the source definition~\cite{thompson1984reflections}.  

\subsection{Real-World Motivations and Benchmark Gaps}
\label{sec:real-world}

\noindent
\paragraph{Supply-Chain Concerns.}
Malicious compilers, third-party model repositories, and hardware-level backdoors each represent potential insertion points for structural attacks~\cite{tehranipoor2010survey}. The \emph{Shadow Logic} proof-of-concept~\cite{hiddenlayer2024} is a vivid example of subgraph-level backdoors slipping past standard checks.

\noindent
\paragraph{Trojanized Models in the Wild.}
Since late 2024, Protect AI’s public \emph{Guardian} scanner has been flagging architectural backdoor patterns (signatures \texttt{PAIT-ONNX-200} and \texttt{PAIT-TF-200}) in open-source ONNX and TensorFlow models hosted on the Hugging Face Hub. A six-month progress report states that Guardian has scanned 4.47 million model versions and raised 352,000 unsafe or suspicious findings across 51,700 models, noting that the new detectors “identified additional architectural backdoors” in both formats~\cite{ProtectAI2025SixMonthReport}. One flagged example is the publicly available \texttt{SoccerTwos.onnx} file, which Guardian labels “Suspicious-ONNX model contains architectural backdoor.” These scanner results show that backdoor-like sub-graphs already appear in real model repositories, although no public analysis has yet demonstrated that the dormant branches can be triggered in practice. We therefore treat them as suspicious architectural backdoors rather than confirmed exploits, while still highlighting their significance for supply-chain security.

\paragraph{LLM‑Generated Hardware Trojans.}
Faruque \emph{et al.}~\cite{Faruque2024GHOST} show that LLMs can assist in generating stealthy hardware Trojans. While not compromising the LLM itself, these findings underscore that large models can be used to design advanced backdoors.

\noindent
\paragraph{Benchmarking Limitations.}
Datasets like \emph{BackdoorBench}~\cite{wu2022backdoorbench} focus mostly on data- or weight-centric backdoors, with limited coverage of structural infiltration and compiler-level vulnerabilities~\cite{pang2023trojanzoo}. This gap hinders validation of new defense strategies on realistically compromised architectures, motivating specialized benchmarks for structural threats.

\section{Taxonomy of Architectural Backdoors}
\label{sec:taxonomy}
Following the taxonomy proposed by Langford \emph{et al.} \cite{Langford2025}, whose user study showed that only 37\% of professional reviewers noticed an injected sub-graph, we categorize architectural backdoor attacks based on how and when the adversary injects malicious functionality. Broadly, we identify four categories: (1) backdoors embedded directly into the model’s architecture design (via specialized trigger structures within the network, as detailed in §\ref{sec:trigger-structures}); (2) backdoors introduced through a compromised build toolchain or compiler (§\ref{sec:comp-attack}); (3) backdoors arising from AutoML processes such as NAS (§\ref{sec:automl_attacks}); and (4) hybrid approaches that combine multiple mechanisms to implant backdoors (§\ref{sec:hybrid-attacks}). We discuss each in turn in the following subsections.

\begin{figure*}[t]
  \centering
  \resizebox{\textwidth}{!}{%
  \begin{tikzpicture}[
      >=Stealth,
      node distance=1.3cm and 1.1cm,
      arr/.style    ={->, thick},
      attacker/.style={draw, fill=red!15, rounded corners,
                       minimum width=2.4cm, minimum height=0.8cm,
                       align=center, font=\small},
      inject/.style  ={draw, rounded corners,
                       minimum width=2.4cm, minimum height=0.8cm,
                       align=center, font=\small},
      stage/.style   ={draw, rounded corners,
                       minimum width=2.4cm, minimum height=0.8cm,
                       align=center, font=\small}
    ]

    % attacker (left, centred vertically on the NAS row)
    \node[attacker] (atk) {Attacker\\(malicious actor)};

    % three insertion points stacked to the right of attacker
    \node[inject, right=of atk, yshift=1.3cm]   (arch){Tainted\\architecture};
    \node[inject, right=of atk]                 (nas) {Compromised\\NAS pipeline};
    \node[inject, right=of atk, yshift=-1.3cm]  (comp){Malicious\\compiler/export};

    % pipeline stages (one horizontal row)
    \node[stage, right=3.2cm of nas]  (train){Victim trains model\\on clean data};
    \node[stage, right=of train]      (deploy){Model deployed};
    \node[stage, right=of deploy]     (trigger){Trigger input};
    \node[stage, right=of trigger]    (exploit){Backdoor fires};

    % attack arrows
    \draw[arr, red] (atk.east) -- (arch.west);
    \draw[arr, red] (atk.east) -- (nas.west);
    \draw[arr, red] (atk.east) -- (comp.west);

    % straight arrows from insertion points to training
    \draw[arr] (arch.east) -- (train.west);
    \draw[arr] (nas.east)  -- (train.west);
    \draw[arr] (comp.east) -- (train.west);

    % downstream pipeline arrows
    \draw[arr] (train)  -- (deploy);
    \draw[arr] (deploy) -- (trigger);
    \draw[arr] (trigger)-- (exploit);
  \end{tikzpicture}}%
  \Description{Flow-chart showing static graph analysis, dynamic profiling, and formal verification modules in the proposed detection pipeline.}
  \caption{\textbf{End‑to‑end threat model for architectural backdoor insertion.}
           A malicious actor can embed a backdoor at three supply‑chain stages: tainted architecture
           code, a compromised NAS pipeline, or a malicious compiler/export tool.  Each attack path
           flows into a seemingly clean training phase; after deployment, the hidden logic is invoked
           by a crafted trigger, causing targeted misbehavior.}
  \label{fig:supplychain-threat}
\Description{Swim-lane diagram showing an attacker injecting a tainted architecture, NAS pipeline, or compiler; the model is then trained on clean data, deployed, and finally misbehaves when a trigger input arrives.}  
\end{figure*}
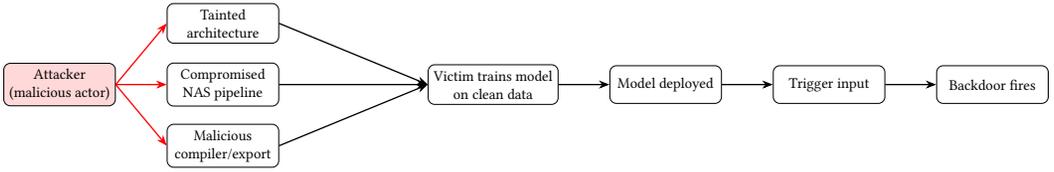

\subsection{Comprehensive Attack Vectors and Detection Barriers}
\label{sec:comp-attack-vectors}

Table~\ref{tab:taxo-detailed} summarizes four principal architectural backdoor
vectors, listing representative mechanisms, threat models and the challenges each poses for detection.  Attackers may also blend these vectors; for example, a NAS-generated design can be passed through a tainted compiler that grafts an extra stealth branch post-training.  Recent forensic scans confirm this hybrid route: benign architectures exported via tampered ONNX or TensorFlow pipelines acquired dormant sub-graphs only at serialization time \cite{protectai2024PAIT}.

% \begin{table}[htbp]
%   \caption{\textbf{Architectural Backdoor Taxonomy: Mechanisms, Threat Models, and Detection Challenges}}
%   \label{tab:taxonomy_comprehensive}
%   \centering
%   \footnotesize             % ACM’s preferred “small table” size
%   \begin{tabularx}{\textwidth}{Y Y Y Y}
%     \toprule
%     \textbf{Attack Vector} & \textbf{Mechanism} &
%     \textbf{Threat Model} & \textbf{Detection Challenges} \\
%     \midrule
%     Sub-network attacks &
%       Implant hidden branches or gating modules in the source &
%       Malicious or coerced developer &
%       Difficult to isolate hidden routes via weight-based scans; persists even with clean retraining~\cite{Bober-Irizar2023,Dumford2020} \\
%     \midrule
%     Compiler-based backdoors &
%       Inject backdoor ops (e.g., \emph{ImpNet}, rewired graph) during compilation &
%       Attacker controls build/export pipeline &
%       Source code appears benign; final IR quietly includes stealth triggers~\cite{Clifford2024,hiddenlayer2024,thompson1984reflections} \\
%     \midrule
%     AutoML/NAS-based backdoors &
%       Manipulate search objective or seed backdoor sub-blocks into candidate layers &
%       Attacker modifies NAS or AutoML reward function &
%       Complex auto-generated architectures obscure small backdoor subgraphs~\cite{pang2022,pang2023dark} \\
%     \bottomrule
%   \end{tabularx}
%   \Description[Taxonomy table]{Four-column table that lists attack vector, mechanism, threat model, and detection challenges for three classes of architectural backdoors.}
% \end{table}

\subsubsection{Detailed Trigger \& Subgraph Structures}
\label{sec:trigger-structures}

Architectural backdoors span multiple structural dimensions: some hard-code constant patterns, others exploit operator quirks; some rely on isolated branches while others weave logic into shared paths, and their misbehavior may be targeted or untargeted. Although the literature largely centers on architectural backdoors that force targeted misclassification, the prospect of untargeted, disruptive behaviors should not be ignored. To make the discussion concrete, we categorize the ways a trigger can be embedded in a model’s computational graph into three representative patterns:  

\begin{itemize}
  \item \textbf{Single-Layer Trigger}\,:  
        One neuron or micro-layer is modified (or newly inserted) so that it remains dormant on benign inputs yet, upon a specific activation pattern, drives the network toward the attacker’s goal.  Because the effect is concentrated in one place, this is usually the easiest variant to detect and prune.

  \item \textbf{Subgraph Trigger (A2)}\,:  
        A single inserted branch or sub-network is grafted onto the host graph.  The branch may share early activations but, once its local trigger fires, it bypasses or overrides the main forward path. Bober-Irizar \textit{et al.} introduced this pattern with a checkerboard branch that survives full retraining \cite{Bober-Irizar2023}. Because the subgraph often re-uses benign neurons, isolating and excising it is substantially harder than with a single-layer trigger.

  \item \textbf{Distributed / Interleaved Trigger (A3)}\,:  
        Trigger logic is split across multiple layers, heads, or even modalities; no individual component appears malicious in isolation. Only a specific constellation of partial triggers activates the hidden route.  Classic examples include \emph{TrojanNet}, whose keyed weight permutations embed a covert model \cite{Guo2021TrojanNet}, and the \emph{Set / Get / Steer} batch-leakage gates of Küchler \textit{et al.} \cite{Kuchler2025BatchSteal} where Set overwrites a victim's output; Get leaks it; Steer subtly biases another user’s result—all without direct access. Beyond vision and NLP, Chen \textit{et al.} demonstrate a relation-aware trigger for heterogeneous graph neural networks (HGBA) that inserts only a handful of edges yet achieves near-perfect attack success \cite{Chen2025HGBA}. Because every fragment piggy-backs on normal activations, graph-diff and
        activation-clustering defenses frequently fail \cite{Chen2019ActivationClustering}.
\end{itemize}

%%%%%%%%%%%%%%%%%%%%%%%%%%%%%%%%%%%%%%%%%%%%%%%%
% Figure 5: Separate-Path Backdoor (east–then–down)
%%%%%%%%%%%%%%%%%%%%%%%%%%%%%%%%%%%%%%%%%%%%%%%%
\begin{figure}[htbp]
\centering
\begin{tikzpicture}[
    node distance = 1.2cm and 1.5cm,
    box/.style    = {draw, rectangle, minimum height=0.4cm,
                     minimum width=2.2cm, font=\footnotesize},
    trigger/.style = {draw, rectangle, fill=red!20, minimum height=0.4cm,
                     minimum width=2.6cm, font=\footnotesize},
    ->, >=Stealth
]
% Main path
\node[box] (s_input)  {Input};
\node[box, right=of s_input]  (s_main)  {Main Path};
\node[box, right=of s_main]   (s_output){Output};

% Separate malicious branch
\node[trigger, above=0.4cm of s_main] (s_troj) {Backdoor Path};

% --- Edges -------------------------------------------------
\draw[->] (s_input) -- (s_main) -- (s_output);            % normal flow

% malicious flow (thick red dashed lines)
\draw[->, red, dashed, line width=1pt] (s_input.north) |- (s_troj.west);  % up then right
\draw[->, red, dashed, line width=1pt] (s_troj.east)  -| (s_output.north);% right then down
% ----------------------------------------------------------

\end{tikzpicture}
\caption{\textbf{Separate-path backdoor.} An isolated malicious circuit (red dashed path) bypasses normal computation entirely and forces the output once its trigger fires.}
\label{fig:sep_path}
\Description{Network graph with a thick red dashed branch departing from the input, passing through a hidden backdoor module, then moving east and down to the center top of the output node.}
\end{figure}
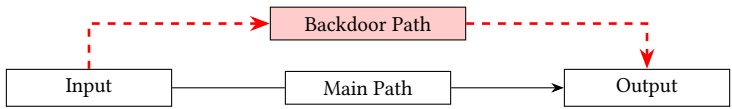

%%%%%%%%%%%%%%%%%%%%%%%%%%%%%%%%%%%%%%%%%%%%%%%%
% Figure 6: Shared-Path Backdoor (tall box, text below arrow)
%%%%%%%%%%%%%%%%%%%%%%%%%%%%%%%%%%%%%%%%%%%%%%%%
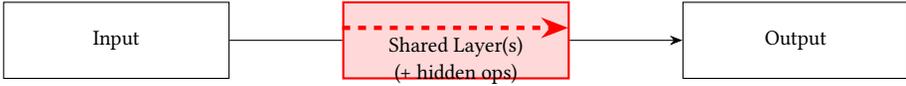
\begin{figure}[htbp]
\centering
\begin{tikzpicture}[
    node distance = 1.2cm and 1.5cm,
    box/.style     = {draw, rectangle, minimum height=1.0cm,   % 1 cm tall
                      minimum width=3cm, font=\footnotesize},
    cleanarrow/.style ={->},
    badsignal/.style  ={->, red, dashed, ultra thick},
    >=Stealth
]
% Nodes ---------------------------------------------------
\node[box] (input) {Input};

% Shared layer: tall, red outline + light fill
\node[box, draw=red, thick, fill=red!15,
      right=of input] (shared) {};

\node[box, right=of shared] (output) {Output};

% --------------------------------------------------------
% Put text node inside shared layer, anchored at bottom
\node[font=\footnotesize, align=center] at ($(shared.south)+(0,0.25)$)
      {Shared Layer(s)\\(+ hidden ops)};

% Normal benign flow -------------------------------------
\draw[cleanarrow] (input) -- (shared) -- (output);

% Bold covert arrow near the top edge --------------------
\draw[badsignal]
      ($(shared.north west)+(0,-0.35)$) --
      ($(shared.north east)+(-0.02,-0.35)$);

\end{tikzpicture}
\caption{\textbf{Shared-path backdoor.} Malicious computations reside inside the shared layer; the bold red dashed arrow across the top shows the hidden signal, while legitimate processing and label text sit safely below it.}
\label{fig:shared_path}
\Description{A tall red-outlined layer contains the hidden backdoor arrow near its top, with descriptive text centered lower in the box and away from the arrow.}
\end{figure}

%%%%%%%%%%%%%%%%%%%%%%%%%%%%%%%%%%%%%%%%%%%%%%%%
% Figure 7: Interleaved-Path Backdoor (revised)
%%%%%%%%%%%%%%%%%%%%%%%%%%%%%%%%%%%%%%%%%%%%%%%%
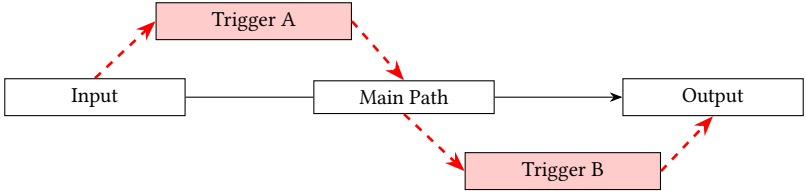
\begin{figure}[htbp]
\centering
\begin{tikzpicture}[
    node distance = 1.2cm and 1.7cm,
    box/.style     = {draw, rectangle, minimum height=0.42cm,
                      minimum width=2.4cm, font=\footnotesize},
    trigger/.style = {draw, rectangle, fill=red!20, minimum height=0.42cm,
                      minimum width=2.6cm, font=\footnotesize},
    ->, >=Stealth
]

% -----------------------------------------------------------------
% Main benign path
% -----------------------------------------------------------------
\node[box]                     (i_input)  {Input};
\node[box, right=of i_input]   (i_main)   {Main Path};
\node[box, right=of i_main]    (i_output) {Output};

\draw[->] (i_input) -- (i_main) -- (i_output);

% -----------------------------------------------------------------
% Partial trigger detectors (distributed)
% -----------------------------------------------------------------
\node[trigger, above right=0.5cm and -0.4cm of i_input] (i_pt1) {Trigger A};
\node[trigger, below right=0.5cm and -0.4cm of i_main]  (i_pt2) {Trigger B};

% -----------------------------------------------------------------
% Interleaved malicious signal  (thick red dashed)
%   • exits main path → Trigger A
%   • rejoins at i_main.north
%   • exits again → Trigger B
%   • final merge at i_output.south
% -----------------------------------------------------------------
\draw[->, red, dashed, line width=1pt] (i_input.north) -- (i_pt1.west);
\draw[->, red, dashed, line width=1pt] (i_pt1.east)         -- (i_main.north);
\draw[->, red, dashed, line width=1pt] (i_main.south)       -- (i_pt2.west);
\draw[->, red, dashed, line width=1pt] (i_pt2.east)         -- (i_output.south);

\end{tikzpicture}
\caption{\textbf{Interleaved-path backdoor.} A hidden signal (red dashed line) repeatedly diverges from and rejoins the main computation path, with each hop gated by a different partial trigger.  All fragments must activate to corrupt the output.}
\label{fig:interleaved_path}
\Description{Diagram with an Input box, a Main Path box, and an Output box in a line. Above Input sits Trigger A; below Main Path sits Trigger B. A thick red dashed arrow leaves the main path to Trigger A, rejoins the main path, leaves again to Trigger B, and finally merges into the Output node.}
\end{figure}

%%%%%%%%%%%%%%%%%%%%%%%%%%%%%%%%%%%%%%%%%%%%%%%%
% Figure 8: NAS-Based Backdoor (revised)
%%%%%%%%%%%%%%%%%%%%%%%%%%%%%%%%%%%%%%%%%%%%%%%%
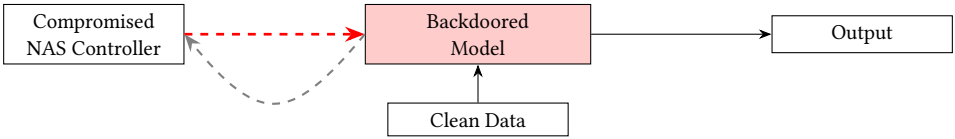
\begin{figure}[htbp]
\centering
\begin{tikzpicture}[
    font=\footnotesize,
    node distance = 2cm and 2.4cm,
    box/.style      = {draw, rectangle, minimum height=0.42cm,
                       minimum width=2.4cm, align=center},
    malicious/.style= {draw, rectangle, fill=red!20, minimum height=0.42cm,
                       minimum width=3.0cm, align=center},
    ->, >=Stealth
]
% -----------------------------------------------------------------
% (1) Compromised NAS controller
% -----------------------------------------------------------------
\node[box] (controller) {Compromised\\NAS Controller};

% -----------------------------------------------------------------
% (2) Backdoored model produced by NAS
% -----------------------------------------------------------------
\node[malicious, right=of controller] (model) {Backdoored\\Model};

% -----------------------------------------------------------------
% (3) Clean training / inference data
% -----------------------------------------------------------------
\node[box, below=0.5cm of model] (input) {Clean Data};

% -----------------------------------------------------------------
% (4) Normal output head
% -----------------------------------------------------------------
\node[box, right=of model] (output) {Output};

% -----------------------------------------------------------------
% Edges
% -----------------------------------------------------------------
% Malicious design edge
\draw[->, red, dashed, line width=1pt] (controller) -- (model);

% Optional: show NAS search loop (dashed grey)
\draw[->, dashed, thick, gray]
      (model.west) .. controls +(-1.0,-1.2) and +(1.0,-1.2) .. (controller.east);

% Clean training/inference flow
\draw[->] (input) -- (model.south);
\draw[->] (model.east) -- (output.west);

\end{tikzpicture}
\caption{\textbf{NAS-based backdoor.} A compromised AutoML (NAS) controller emits a backdoored model (red), even when trained on clean data.  At inference time, the model behaves normally, unless the hidden trigger is present.}
\label{fig:nas_backdoor}
\Description{Diagram with a “Compromised NAS Controller” feeding a “Backdoored Model” via a red dashed arrow; a gray dashed arrow loops back to the controller, indicating the search process. Clean data enters the model from below, and a normal black arrow leads to an Output node.}
\end{figure}

\begin{table}[htbp]
  \caption{\textbf{Expanded 12-Category Taxonomy (Adapted from \cite{Langford2025}).}
  \label{tab:taxo-detailed}
  We break each of the four overarching categories, (A) Sub-network attacks,
  (B) Compiler-based backdoors, (C) AutoML/NAS-based backdoors, and
  (D) Hybrid attacks, into three subcategories, yielding 12 distinct attack
  vectors. Each row summarizes the mechanism, threat model, and detection
  challenges, referencing relevant prior work.}
  \label{tab:taxonomy_table9}
  \centering
  \scriptsize           
  \begin{tabular*}{\columnwidth}{@{\extracolsep{\fill}}
      P{2.4cm} P{4.0cm} P{2.8cm} P{3.7cm}}
    \toprule
    \textbf{Subcategory} & \textbf{Mechanism} & \textbf{Threat Model}
      & \textbf{Detection Challenges} \\
    \midrule

    \textbf{(A1) Single-Layer Trigger} &
    A single neuron or layer remains dormant except on a specific pattern,
    causing misclassification. &
    Malicious architect or insider with direct control over model design. &
    The local trigger can be subtle yet persistent, evading naive weight
    inspection \cite{Bober-Irizar2023}. \\ \midrule

    \textbf{(A2) Subgraph Trigger} &
    A small subnetwork is embedded for the backdoor, sometimes partially
    used in normal computation. &
    Attacker modifies the official architecture code or design blueprint. &
    Because a backdoor sub-network can reuse neurons that also serve normal computation, identifying its malicious role is substantially harder than for a single rogue layer~\cite{Bober-Irizar2023}. \\ \midrule

\textbf{(A3) Distributed / Interleaved Trigger} &
Trigger logic is spread across multiple layers or nodes, each benign in isolation.  Examples range from TrojanNet’s hidden-weight permutation model~\cite{Guo2021TrojanNet} to the recent batch-context leakage gate that copies features across examples within a single inference batch~\cite{Kuchler2025BatchSteal}. &
Attacker designs complex network topologies with multiple partial triggers, structural switches, or cross-example paths. &
Requires holistic graph analysis to detect; no individual element exhibits an obvious malicious pattern \cite{Guo2021TrojanNet}. \\ \midrule

    \textbf{(B1) Malicious IR-Rewriting} &
    \emph{ImpNet} inserts a hidden branch directly into the IR during compilation. &
    Attacker controls the build/export pipeline (cf.\ Thompson’s “trusting-trust”). &
    Source code looks clean; the injected branch is visible only in the compiled graph
    \cite{Clifford2024,thompson1984reflections}. \\ \midrule

    \textbf{(B2) Compiler-Level Operator Injections} &
    Standard ops (Conv, ReLU, etc.) replaced with modified versions that
    hide triggers. &
    Root-level access to compiler or operator libraries. &
    Subtle operator changes do not alter normal accuracy; triggers remain
    dormant until specific input \cite{hiddenlayer2024}. \\ \midrule

    \textbf{(B3) Trojaned Export/ Serialization} &
    A malicious export tool adds extra nodes or branches when converting to
    ONNX/TFLite. &
    Attacker can tamper with final serialization routines post-training. &
    The “deployed” model differs from the validated training model, rarely re-checked in practice \cite{hiddenlayer2024,protectai2024PAIT}. \\ \midrule

    \textbf{(C1) Malicious Reward Shaping in NAS} &
    NAS objective or reward is altered so that backdoor-friendly designs get
    higher scores. &
    Attacker modifies the AutoML pipeline to favor certain sub-blocks or
    triggers. &
    Models appear optimal on clean data, yet contain dormant subgraphs by
    design \cite{pang2022}. \\ \midrule

    \textbf{(C2) Poisoned Search Space} &
    Malicious building blocks or candidate layers in the search space embed
    dormant triggers. &
    Third-party or coerced library contributor seeds hidden modules. &
    The hidden module is simply “selected” by the NAS algorithm; easy to
    miss unless the search space is audited \cite{pang2023dark}. \\ \midrule

    \textbf{(C3) Injected Sub-Block Shortcuts} &
    Small sub-blocks with conditional shortcuts are inserted in the final
    architecture. &
    Compromised AutoML code that automatically includes the sub-block in
    each candidate design. &
    Disguised as normal architectural blocks, they only activate under a
    special input pattern \cite{pang2022}. \\ \midrule

    \textbf{(D1) Architecture + Compiler Synergy} &
    Adversary plants a structural backdoor, then obfuscates it further with
    malicious compile steps. &
    Full pipeline control (model design + compiler). &
    Redundant triggers: removing one layer of the backdoor might leave
    another intact \cite{Clifford2024}. \\ \midrule

    \textbf{(D2) Architecture + Data Poisoning} &
    Mild data poisoning supports a hardware-level or structural backdoor. &
    Attacker can tamper with both training data and the architecture. &
    Defenses focusing only on data anomalies or only on structural audits
    can miss the combined effect \cite{pang2023dark}. \\ \midrule

    \textbf{(D3) Multi-Stage Obfuscation} &
    Multi-phase approach: malicious design + data poisoning + compile-level
    changes. &
    Well-resourced attacker infiltrating every stage of the model lifecycle. &
    Each malicious layer is benign on its own; the final deployed model
    exhibits stealthy, robust triggers \cite{Bober-Irizar2023}. \\
    \bottomrule
  \end{tabular*}
  %  Accessible summary for HTML conversion (ACM requirement)
  \Description[12-category attack taxonomy table]{Table summarizing mechanism,
  threat model, and detection challenges for twelve architectural backdoor
  attack vectors across sub-network, compiler-based, AutoML/NAS-based, and
  hybrid categories.}
\end{table}

\subsubsection{Compiler-Based Backdoors}
\label{sec:comp-attack}

Modern deep-learning pipelines rely on compilers and code-generation tools (e.g., ONNX converters or mobile-deployment frameworks) to translate high-level model definitions into optimized executables.  An attacker who compromises this toolchain can inject hidden components during compilation, exactly what \textit{ImpNet} demonstrates by inserting a weight-independent branch that remains invisible to source-level audits \cite{Clifford2024}. The technique is the modern analogue of Thompson’s classic compiler trojan \cite{thompson1984reflections}: the compiler surreptitiously adds extra neurons, connections, or conditions that implement a dormant trigger without the developer’s knowledge. More recnt work showed 269 real-world ONNX models had data leakage due to poor operator design \cite{Kuchler2025BatchSteal}. The leakage was traced to misuse of the \texttt{DynamicQuantizeLinear} operator, which improperly shared state across batch entries
\noindent
Recent findings by Protect AI provide the first confirmed examples of such compiler-level insertions in the wild~\cite{protectai2024PAIT}. Their \texttt{PAIT-ONNX-200} and \texttt{PAIT-TF-200} disclosures document backdoored models uploaded to public hubs where hidden branches were introduced during ONNX export or TensorFlow SavedModel serialization. These subgraphs, invisible at the source level, lay dormant until specific input triggers were received, demonstrating the feasibility of real-world IR-level manipulation and silent structural sabotage.
\noindent
In a related threat model, Zhu~\textit{et al.}~\cite{Zhu2024TFMalware} show that TensorFlow's export APIs can be abused to embed executable logic, such as shell commands, directly into SavedModel graphs. These logic bombs activate on deserialization, turning AI models into malware containers without affecting model predictions, further expanding the architectural attack surface at serialization time.
\noindent
Qi~\textit{et~al.} extend this idea to the deployment stage: an adversary can overwrite only a handful of filters to splice a one-channel hidden subgraph into a trained CNN, achieving $>99\%$ attack success rates (ASR) on \emph{ImageNet} while reducing clean accuracy by less than 2\%, all without training data or gradients~\cite{Qi2023SRA}.

\subsubsection{AutoML/NAS-Based Backdoors}
\label{sec:automl_attacks}
Beyond manual design and compiler manipulation, recent work has revealed that AutoML processes themselves can introduce backdoors. AutoML pipelines, especially NAS, can be deliberately manipulated to yield models with hidden architectural backdoors. Pang \textit{et al.}~\cite{pang2022,pang2023dark} explicitly introduce the exploitable and vulnerable
arch search \emph{(EVAS)} and show that, by adding a malicious reward term, an adversary can steer NAS to select architectures containing a dormant “shortcut’’ subgraph which activates on a specific trigger pattern. 
\noindent
This AutoML-based backdoor attack is especially insidious because it does not require any explicit tampering with training data or model parameters. What survives is the capacity for malicious behavior: the hard-wired branch remains present, and standard training loss will quickly re-learn suitable weights, so the attacker needs little or no additional effort to reactivate the payload. Moreover, since the compromised model’s weights and training data may appear entirely benign, such architecture-level backdoors naturally evade many conventional defenses that focus on detecting anomalies in the training process or parameter values. This emerging class of attacks highlights a new security risk in AutoML pipelines, calling for heightened scrutiny of NAS-generated models. Although the Protect AI incidents involved post-training insertion, they underscore that even cleanly designed NAS-generated architectures can be compromised during export, highlighting the need for end-to-end validation of AutoML pipelines.

\subsubsection{Hybrid and Combined Modes}
\label{sec:hybrid-attacks}

Hybrid or multi-stage backdoor attacks could pose a particularly potent threat because they can evade defenses designed to detect single-stage insertions. For instance, an attacker might first leverage NAS to introduce subtle vulnerabilities during architecture search and subsequently manipulate the compiler or model-export phase to further conceal or strengthen the malicious logic~\cite{pang2023dark,Clifford2024}. Such multi-stage approaches reflect real-world attack sophistication, requiring defenders to adopt equally comprehensive detection and mitigation strategies.
\noindent
Attackers can combine methods, for instance feeding a NAS-generated graph through a tainted compiler that grafts a stealth branch while also performing mild data poisoning during training. Such hybrid backdoors blend architectural and training-time vectors, allowing one component to survive even if the other is mitigated. Because these multi-stage exploits bypass defences that target a single insertion point, effective protection must monitor the entire model life-cycle.

\subsection[Expanded 12-Category Taxonomy]{Expanded 12-Category Taxonomy (Adapted from \cite{Langford2025})}
Building on the twelve-subcategory framework of Langford\,\emph{et al.} \cite{Langford2025}, we further break each family into three sub-types, yielding twelve distinct architectural attack vectors. 
\noindent
Table~\ref{tab:taxonomy_table9} summarizes these subcategories along with their mechanisms, threat models, and detection challenges \cite{Bober-Irizar2023, Dumford2020, pang2022, pang2023dark, Clifford2024, thompson1984reflections}.

\section{Detection of Architectural Backdoors}
\label{sec:detection}

Architectural backdoors often resist established detection methods, which typically focus on suspect weight distributions or poisoned training samples, because topology-level backdoors evade traditional anomaly checks. For instance, activation clustering ~\cite{Chen2019ActivationClustering} or spectral signatures ~\cite{tran2018spectral} detect unusual neuron activations triggered by poisoned data, but an architectural backdoor can remain dormant on clean data, producing no detectable activation anomalies. Similarly, methods like \emph{Neural Cleanse}~\cite{Wang2019}, designed to reverse-engineer visible single-input triggers, may fail to detect architectural triggers that depend on internal multi-branch conditions or composite input patterns. Recent research into “defense-aware” attacks specifically aims to evade such inversion methods~\cite{Miah2024ArchLLM}. Thus, specialized detection strategies for architectural threats are critically needed. 
\noindent
In this section, we provide:

\begin{itemize}
    \item Key Definitions and Motivations: Explaining why topology-level backdoors evade traditional anomaly checks.
    \item Structured Detection Approaches: Grouped into static graph inspection, dynamic/trigger inversion, explainability/\allowbreak meta-analysis, and formal verification methods.
    \item Challenges, Benchmarks, and Metrics: Highlighting the real-world obstacles to detection and the limited coverage of existing benchmarks.
\end{itemize}

% TODO: merge some of this with definitions subsection of Background section
% \subsection{Key Definitions}
% \textbf{Trigger} – any input pattern or condition that, when recognized by a malicious subgraph, forces the network to produce an attacker-defined output. Triggers may appear as small pixel patches \cite{Gu2017badnets}, token sequences, or subtle spectral changes \cite{liu2025ladder} and are typically designed to remain inconspicuous on standard tests.

% \noindent
% \textbf{Malicious subgraph} – a structural addition or alteration within the computational graph that, upon detecting the trigger, overrides the standard inference path. It often remains dormant for clean inputs, evading detection through conventional parameter-level analysis \cite{Bober-Irizar2023}.

% \noindent
% \textbf{Compiler‑level insertion} – Although attackers may modify the original source, an equally serious threat arises from tampering at the \emph{compilation} or \emph{export} phase. In this scenario, a backdoor subgraph is introduced post-training by a compromised build tool; these hidden instructions may reside solely in the final model artifact \cite{Clifford2024}.
%\smallskip

\begin{figure}[t]
\centering
\begin{tikzpicture}[
    font=\footnotesize,
    box/.style={
        draw,
        rectangle,
        rounded corners,
        fill=blue!10,
        minimum width=2.3cm,
        minimum height=0.5cm,
        align=center
    },
    stage/.style={
        draw,
        rectangle,
        rounded corners,
        fill=green!15,
        minimum width=2.3cm,
        minimum height=0.5cm,
        align=center
    },
    thickarrow/.style={->, >=latex, thick},
    every node/.style={align=center}
]
% Bottom node: Final Model
\node[box] (model) at (0,0) {Final Model\\(Trained/Compiled)};

% Four method nodes above the model
\node[stage] (static)  at (-4, 2) {Static Graph\\Inspection};
\node[stage] (dynamic) at (-1.3, 2) {Dynamic\\Probing};
\node[stage] (explain) at (1.3, 2) {Explainability\\ \& Meta-Analysis};
\node[stage, fill=yellow!20] (formal) at (4,2) {(Semi-)Formal\\Verification};

% Arrows from the Final Model to each method
\draw[thickarrow] (model) -> (static);
\draw[thickarrow] (model) -> (dynamic);
\draw[thickarrow] (model) -> (explain);
\draw[thickarrow] (model) -> (formal);
\end{tikzpicture}
\caption{\label{fig:detection-overview}
\textbf{Overall detection pipeline.} Defenders can apply one or more methods (static, dynamic, explainability, and (Semi-) formal
verification) to reveal suspicious subgraphs.}
\Description{Stacked schematic: the compiled model sits at the bottom; above it, colored blocks represent static graph diffing, dynamic probing, explainability analysis, and semi-formal verification.} 
\end{figure}
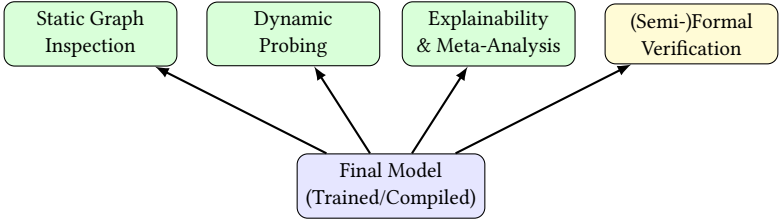

\noindent
Architectural backdoors are tackled via four complementary families: (i) \emph{static graph inspection}~(§\ref{sec:static-graph}), (ii) \emph{dynamic trigger discovery}~(§\ref{sec:dynamic-probing}), (iii) \emph{explainability \& meta‑analysis}~(§\ref{sec:Explainability-Based-Detection}), and (iv) \emph{(semi‑)formal verification}~(§\ref{sec:formal-verification}). Each family trades scalability against completeness, so a layered pipeline (Fig.~\ref{fig:detection-overview}) is recommended in practice.

\subsection{Static Graph Analysis and Model Introspection}
\label{sec:static-graph}

\begin{figure}[t]
\centering
\begin{tikzpicture}[
  font=\footnotesize,
  node distance = 1.7cm and 2.2cm,
  box/.style      ={draw, rectangle, rounded corners,
                    fill=blue!10,
                    minimum width=2.6cm, minimum height=0.5cm, align=center},
  highlight/.style={draw=red, thick, rectangle, rounded corners,
                    fill=red!15,
                    minimum width=2.6cm, minimum height=0.5cm, align=center},
  cleana/.style   ={->, thick, >=Stealth},
  badarrow/.style ={->, red, dashed, ultra thick, >=Stealth}
]

% -------------------------------------------------------------------
% Nodes
\node[box] (source) {Source\\Model Graph};

\node[highlight, right=2.5cm of source] (compiler)
      {Malicious\\Compiler /\\Export Tool};

\node[highlight, below=0.5cm of compiler] (backdoor)
      {Embedded\\\textcolor{red}{Malicious Subgraph}};

\node[box, right=2.5cm of compiler] (binary) {Compiled\\Artifact};

\node[box, below=1.5cm of binary] (diff) {Graph Diff\\Check};

% ------------------------------------------------------------------
% Legitimate flow arrows (black)
\draw[cleana] (source.east)   -- (compiler.west)
      node[midway, above]{compilation};

\draw[cleana] (compiler.east) -- (binary.west)
      node[midway, above]{final model};

\draw[cleana] (binary.south)  -- (diff.north)
      node[midway, right]{re-export};

\draw[cleana] (source.south)  |- (diff.west)
      node[pos=0.7, above]{reference};

% ------------------------------------------------------------------
% Malicious insertion path (bold red dashed)
\draw[badarrow] (compiler.south) -- (backdoor.north);

\draw[badarrow]
      (backdoor.east) -| +(0.4,0) |- (binary.west)
      node[pos=0.25, right]{\scriptsize malicious insert};

\end{tikzpicture}

\caption{\textbf{Compiler-level insertion and static inspection.}  
A compromised compiler injects a hidden subgraph into the compiled artifact.  
A subsequent \emph{graph-diff} step detects the discrepancy by comparing the reference model graph with the potentially compromised artifact.}
\label{fig:compiler_diff}
\Description{Flowchart showing source model, malicious compiler that inserts a red subgraph, compiled artifact, and a graph-diff check that detects the extra nodes.}
\end{figure}
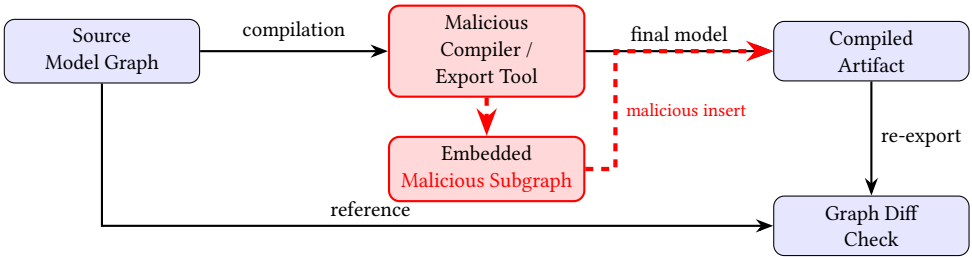

Static graph analysis inspects the exported model (ONNX, TorchScript, etc.) for sub‑paths or operators that diverge from the intended architecture. Architectural backdoors typically surface as extra gating layers, direct input‑to‑output bypasses, or custom ops absent from the high‑level source \cite{Bober-Irizar2023,pang2022}. Figure~\ref{fig:compiler_diff} illustrates how a clean source can compile into a binary with an implanted subgraph. Practically, integrating static analyzers that systematically compare exported model formats (e.g., ONNX or TensorFlow GraphDef) against the original intended architecture can significantly streamline anomaly detection. Static inspection is lightweight and flags obvious structural backdoors, but it often misses obfuscated or deeply-interleaved logic. For example, automated checks for additional layers, unexpected gating logic, or custom operations not defined in the original architecture can be implemented. Maintaining accurate original architecture specifications as reference artifacts during model development is recommended as a best practice for facilitating such comparisons.

% \begin{table}[htbp]
%   \caption{\textbf{Strengths and limitations of static inspection}}
%   \label{tab:static-pros-cons}
%   \centering
%   \footnotesize                                 % ACM-approved small size
%   %
%   % Two ragged-right columns that together fill \columnwidth
%   \begin{tabular*}{\columnwidth}{@{\extracolsep{\fill}}
%           P{0.48\columnwidth} P{0.48\columnwidth}}
%     \toprule
%     \textbf{Strengths} & \textbf{Limitations}\\
%     \midrule
%     % --- single row with two fully populated cells --------------------------
%     \begin{itemize}[leftmargin=*,labelsep=0.4em]
%       \item Lightweight and efficient%
%       \item Detects obvious structural manipulations, gating logic, or
%             custom operations%
%       \item No requirement for training data%
%     \end{itemize}
%     &
%     \begin{itemize}[leftmargin=*,labelsep=0.4em]
%       \item Insensitive to subtle, distributed manipulations or
%             compiler-time insertions%
%       \item May fail to detect dormant or conditionally activated logic%
%       \item Reliant on the availability of trusted reference models%
%     \end{itemize}
%     \\ % ---------------------------------------------------------------------
%     \bottomrule
%   \end{tabular*}
%   \Description[Static-inspection strengths and limitations]{Table listing the
%     strengths (lightweight, detects obvious structure, no data needed) and the
%     limitations (misses subtle or compiler-time insertions, dormant logic,
%     need for trusted reference) of static inspection.}
% \end{table}

\paragraph{Limitations of pruning and purification-based defenses.}
Classical defenses such as fine-pruning \cite{Liu2018FinePruning} aim to remove dormant neurons by eliminating those with low activation on clean data, assuming backdoor logic is encoded in rare weight activations. While effective against weight-based backdoors, this strategy is poorly suited to architectural backdoors, where the malicious logic may reside in the topology itself, e.g., in an extra branch or gated path that is never activated on clean inputs. Recent work further reveals the brittleness of purification-based methods: Gradient Tuning Backdoor Attack++ (\emph{GTBA++}) \cite{min2024gtba} demonstrates that models, even after pruning or adversarial unlearning, can rapidly reacquire high attack success rates when exposed to limited poisoned data. Similarly, the ASR-Proof evaluation framework~\cite{min2024asrproof} demonstrates that many existing defenses only superficially reduce  ASR, failing to eliminate the latent backdoor mechanism. Consequently, attackers can quickly reactivate the backdoor using carefully constructed adaptive queries. These results underscore the need for architecture-aware defences that explicitly reason about structural logic and supply-chain integrity.
\noindent
Static analysis is, therefore, an expedient first line of defense. However, high‑value deployments should complement it with the dynamic probing techniques of §\ref{sec:dynamic-probing}, which can activate stealthy paths that static diffs overlook.

% =============================================================
\subsection{Dynamic Probing and Trigger Inversion}
\label{sec:dynamic-probing}

Dynamic techniques actively search for inputs that route execution through a hidden subgraph. Unlike static diffing (§\ref{sec:static-graph}), they treat the model as a black box and watch for abrupt confidence shifts or label flips. Dynamic probing complements static graph inspection by attempting to uncover stealthy backdoor logic through targeted perturbation of inputs or reverse-engineering potential triggers from model responses. While static approaches analyze the model structure directly, dynamic methods provide evidence through empirical behavior.
\paragraph{Fuzzing.} This randomized approach perturbs model inputs and monitors outputs for unexpected spikes in softmax entropy or abrupt shifts in predicted class probabilities~\cite{rance2022}. Recent fuzzing-based methods, such as Runtime Oracle-guided Search for Backdoor Analysis \emph{(ROSA)}~\cite{marcozzi2023rosa}, use guided fuzzing to discover backdoor triggers in traditional software. Although primarily developed for software vulnerability discovery, analogous fuzzing techniques are gaining attention for ML backdoor detection. Nonetheless, exhaustive fuzzing quickly becomes computationally prohibitive, particularly for complex, composite triggers, highlighting the need for advancements towards multi-modal fuzzing strategies.

\paragraph{Trigger Inversion.} \emph{Neural Cleanse}~\cite{Wang2019} and its extensions~\cite{zeng2022} frame backdoor discovery as an optimization problem, aiming to identify the smallest perturbation necessary to consistently produce targeted misclassification. These methods initially proved effective for single-input triggers, especially in image classifiers, but encounter significant difficulties with multi-branch or composite triggers. Recent research into “defense-aware” backdoors, such as those demonstrated by Miah \emph{et al.}~\cite{Miah2024ArchLLM}, has specifically developed strategies to evade trigger inversion, underscoring the limitations of relying solely on inversion techniques.

\paragraph{Data-Limited Trigger Search.} Recent developments have addressed detection scenarios under harsh practical constraints, notably \textit{DeBackdoor} by Popovic \emph{et al.}~\cite{Popovic2025DeBackdoor}. Operating with only black-box access and extremely limited clean samples (less than $1\%$ of the training dataset), \textit{DeBackdoor} employs a deductive search across an extensive trigger hypothesis space (e.g., shapes, patches, blended patterns). By iteratively optimizing a smoothed attack-success objective through simple forward passes, \textit{DeBackdoor} reconstructs potential triggers without needing large validation sets or direct weight access. Experiments involving multiple architectures, datasets, and attack types demonstrate near-perfect detection accuracy, highlighting its viability for auditing third-party models and uncovering input-dependent manifestations of architectural backdoors. Architectural triggers that alter only logit magnitudes, or otherwise induce minimal output drift, may still dodge current fuzzing and entropy-based heuristics, underscoring the need for more sensitive or adaptive oracles.

\subsection{Explainability–Based Detection and Meta‑Analysis}
\label{sec:Explainability-Based-Detection}

\paragraph{Explainability.}
\emph{SentiNet} \cite{Chou2018Sentinet} applies Gradient-weighted Class Activation Mapping (Grad‑CAM) to locate the most influential regions of an image for a given model decision. Backdoored models often reveal a distinctive heatmap pattern when the trigger is present, as the malicious subgraph sharply concentrates the relevance on the trigger area. Saliency masks or attention weights can thus expose anomalous focus, especially if multiple inputs share a suspiciously consistent activation region. Learning-based meta-detectors such as \emph{MNTD} (Kolouri \emph{et al.}, S\&P 2021) train a binary classifier on model query–response behavior and achieve high AUC in distinguishing Trojaned from clean models without requiring trigger access \cite{Kolouri2021MNTD}.

\paragraph{Meta-analysis}  
\emph{Meta Neural Trojan Detection (MNTD)}~\cite{Kolouri2021MNTD} trains a black-box meta-classifier on a labelled dataset of clean and backdoored models. The classifier then evaluates an unseen model by observing its query–response behavior. A related method is \emph{ABS (Artificial Brain Stimulation)}~\cite{liu2019abs}, which perturbs internal neurons and evaluates outputs to infer latent backdoor presence. While effective on conventional image classifiers, these approaches depend heavily on training coverage and generalization assumptions.
\noindent
Recently, Shen~\textit{et al.} introduced \emph{BAIT} (Backdoor scanning by Inverting the Target)~\cite{Shen2025BAIT}, a meta-level detection method targeting large language models. \emph{BAIT} uses a novel inversion pipeline that traces attacker-desired completions back to hidden trigger patterns embedded in the model’s graph or behavior. Unlike earlier methods that require explicit trigger injections, BAIT reverses from suspicious outputs to uncover latent structural logic, enabling scalable detection of backdoor pathways even in fine-tuned or instruction-tuned LLMs.
\noindent
As LLMs increasingly incorporate adapters and LoRA modules, parameter-efficient fine-tuning also opens new vectors for architectural backdoors. \emph{PEFTGuard}~\cite{Sun2025PEFTGuard} bridges meta-analysis and explainability by inspecting parameter-efficient fine-tuning (PEFT) modules for hidden decision logic and providing visual explanations of their graph-level impact. Although designed for parameter-efficient settings, its structured introspection pipeline is directly relevant to architectural risk modeling in LLM-class systems.
\noindent
Explainability excels when a trigger leaves a spatial or statistical footprint, whereas meta-analysis trades false positives for scalability to unseen models. Newer tools like \emph{BAIT} and \emph{PEFTGuard} signal a promising shift toward inversion-based explainability that generalizes across architectures and formats.

\subsection{Formal Verification and Semi-Formal Methods}
\label{sec:formal-verification}

Formal verification frameworks, such as \emph{Reluplex} \cite{Katz2017Reluplex} and its successor \emph{Marabou} \cite{katz2019marabou} express the backdoor question as: “Is there any input that is almost identical to a normal one yet still flips on the hidden branch?” Here “almost identical” means the input sits inside a very small neighborhood, often called an \textit{$\ell_p$ ball}, around a clean reference example. For images, two common interpretations of that neighborhood are (i) \textbf{per-pixel bound} (no individual pixel is changed by more than a tiny value) and  
(ii) \textbf{overall-energy bound} (the total Euclidean change across all pixels is tiny). If the solver can show that every such near-clone leaves the backdoor branch dormant, the model is certified safe within that radius. If it cannot, the tool returns a concrete offending input, effectively uncovering the trigger~\cite{Pham2022CAV}. A deeper scalability analysis appears in § \ref{sec:future-verif}.

\subsection{Strengths \& Limits of Current Detectors}\label{sec:strengths-limits}

The state of the art in detecting architectural backdoors is still hemmed in by four interlocking hurdles. First, multi‑trigger complexity frustrates current tools: when malicious logic is distributed across several branches or layers it can demand a specific constellation of signals before activation \cite{Langford2025,liu2025ladder, zeng2022}. Second, compiler obfuscation undermines source‑level audits; an apparently clean model definition may compile into an intermediate representation that hides a backdoor subgraph, a strategy dramatized by the \emph{Shadow Logic} proof‑of‑concept \cite{hiddenlayer2024}. Third, both dynamic probing and (semi‑)formal verification carry a steep computational cost once models approach modern scale, so practitioners face a trade‑off between coverage and run‑time analysis.  
Finally, genuine progress is hampered by benchmark scarcity: most public datasets still revolve around pixel‑ or label‑poisoned CNNs, leaving structural exploits to small ad‑hoc collections \cite{pang2022,Bober-Irizar2023,pang2023dark}.
\noindent
Because architectural attacks typically leave clean-set accuracy untouched, the community relies on indicators such as true-positive rate (TPR), false-positive rate (FPR), and attack-success-rate reduction (ASR) achieved after mitigation.  Benchmarks like \emph{BackdoorBench}~\cite{wu2022backdoorbench} and government-sponsored \emph{TrojAI} evaluations~\cite{NISTTrojAI} report these numbers consistently. However, even top detectors in TrojAI primarily demonstrate effectiveness on data-poisoning-based backdoors; these challenges have not extensively tested hidden architectural or compiler-level insertions, leaving a critical evaluation gap for real-world structural threats. An additional community resource is \emph{TrojanZoo}, an open-source corpus of backdoored and clean models released by Pang \emph{et al.} \cite{pang2023trojanzoo}; it provides graph-level ground truth for CNNs and Transformers and is now widely used to sanity-check detection pipelines.
\noindent
A credible next generation should cover at least three scenarios that today’s datasets omit altogether: (i) multi‑branch gating models in which individual paths look benign in isolation; (ii) backdoors introduced only at compile time, visible in the binary graph but absent from the high‑level source; and (iii) transfer‑learning persistence tests that check whether a architectural backdoor survives when the model is fine‑tuned on a new domain \cite{Wang2025TransTroj}. Incorporating these cases would close the evaluation gap and give forthcoming detection methods a realistic proving ground.

% Architectural backdoors are unusually persistent because their logic is inherently structural. Static graph inspection is fast and data-free but may miss compiler-inserted or deeply interleaved paths. Dynamic probing methods (fuzzing, trigger inversion) can uncover hidden routes but struggle with composite or highly obfuscated triggers such as Stealthy Scapegoat Backdoor Attack (SGBA) decoys~\cite{He2024SGBA} or high-frequency keys~\cite{xia2024}. While formal verification offers provable guarantees, it currently scales poorly beyond medium-sized models and norm-bounded triggers. 
\noindent
Table~\ref{tab:detector-comparison} summarizes the complementary strengths and blind
spots of static, dynamic and formal approaches.  Because no single technique is
fool-proof, readers should see the consolidated research gaps in
§\ref{sec:future}.  The next section (§\ref{sec:mitigation}) turns to repair:
how to neutralize flagged sub-graphs, restore benign accuracy, and harden the
pipeline against future inserts.

\begin{table}[htbp]
  \caption{Comparison of static, dynamic and formal detection methods.}
  \label{tab:detector-comparison}
  \centering
  \footnotesize
  \setlength{\tabcolsep}{4pt}
  \begin{tabularx}{\linewidth}{p{2.6cm} Y Y}
    \toprule
    \textbf{Approach} & \textbf{Strengths} & \textbf{Limitations}\\
    \midrule
    Static graph &
      Lightweight; reveals overt gating or custom ops without training data &
      Blind to subtle, distributed, or compiler-time inserts \\ \midrule
    Dynamic probing &
      Confirms malicious behavior on crafted inputs &
      Random/gradient search struggles with multi-condition triggers or
      heavy obfuscation (e.g., SGBA decoys~\cite{He2024SGBA}) \\ \midrule
    Formal verification &
      Provable absence of backdoors within bounded input domain; yields
      counter-example if proof fails &
      Currently unscalable to large, multi-branch models; misses
      spectral/composite triggers; invalidated by source/IR drift \\ \midrule
    Hybrid pipeline &
      Layers static, dynamic and partial proofs for broader coverage &
      Requires orchestration across tools; no standard benchmark for
      supply-chain scale yet \\
    \bottomrule
  \end{tabularx}
\end{table}

\section{Mitigation and Model Repair}
\label{sec:mitigation}

Architectural backdoors (§\ref{sec:detection}) resist naive mitigation because the malicious logic is structurally “wired” into the model's topology. Unlike weight backdoors that one can ‘forget’ by fine-pruning or retraining, a structural backdoor must be surgically removed from the network architecture. This process is complicated by multi-branch logic that scatters gates across layers \cite{pang2022}, topological embedding that survives weight resets, rare trigger activations that evade fine-tuning gradients, and the persistent risk of re-insertion via compromised supply chains. This section surveys practical, graph-level, and supply-chain measures tailored to excise or disable these sophisticated structural threats, with particular emphasis on multi-branch and compiler-time attacks.

\subsection{Subgraph Pruning and Removal}
\label{sec:mitigation-pruning}

Static diffing (§\ref{sec:static-graph}) or dynamic probing (§\ref{sec:dynamic-probing}) first localize a suspect path, potentially using guided fuzzers such as ROSA \cite{marcozzi2023rosa}, then excise that subgraph and optionally fine‑tune to recover accuracy. Classical fine‑pruning of neurons\cite{Liu2018FinePruning} or dataset filtering alone leave the graph intact and thus fail against topological attacks.
\noindent
Bober-Irizar \emph{et al.} report that excising the checkerboard branch slashed ASR from 100 \% to 2 \% (\-98 \%) while leaving clean accuracy intact \cite{Bober-Irizar2023} We suggest practitioners first use static diff (§\ref{sec:static-graph}) or dynamic probing (§\ref{sec:dynamic-probing}) to identify suspicious subgraphs, excise them, and immediately re-check model accuracy. If attack success remains partially elevated (e.g., around 40\%), it indicates additional hidden branches still exist. Practitioners must then iteratively remove these residual branches until attack success is fully neutralized.
\noindent
Complementing these approaches, recent work by Bajcsy and Bros \cite{bajcsy2024sim} introduces a web-based simulation playground that lets practitioners plant, trigger, and defend against cryptographic, including checksum based, architectural backdoors. The sandbox supports realistic plant-and-defend cycles, enabling researchers to validate and stress-test backdoor detection and proximity-analysis defenses.

\begin{table}[htbp]
  \caption{\textbf{Subgraph removal at a glance}}
  \label{tab:pruning-pros-cons}
  \centering
  \footnotesize
  % one fixed-width column + two flexible ragged columns
  \begin{tabularx}{\linewidth}{P{2.7cm} Y Y}
    \toprule
    & \textbf{Strengths} & \textbf{Caveats} \\
    \midrule
    Excise single branch &
      Directly disables known route; no full retrain needed &
      Accuracy loss if branch intertwines with benign features \\ \midrule
    Multi-branch pruning &
      Cuts distributed triggers when combined with iterative search &
      Residual gates survive if any branch escapes detection \\ \midrule
    Compiler-aware pruning &
      Checks IR after build to confirm removal &
      Fails if toolchain is still compromised; must repeat per build \\
    \bottomrule
  \end{tabularx}
\end{table}

\noindent
A hybrid mitigation pipeline should therefore:  
(1) detect all suspect paths;  
(2) prune or redirect them; and  
(3) re‑export the model under a trusted, signed compiler to
prevent silent reinsertion.  Subsequent sections cover adversarial
unlearning and supply‑chain hardening that complement graph excision.

\subsection{Adversarial Unlearning and Re‑Training}
\label{sec:mitigation-retraining}

Once a trigger or suspect gate is located (§\ref{sec:detection}), adversarial unlearning seeks to purge its influence by re‑training the model on that trigger with the correct label. Early work such as \emph{Neural Cleanse} searches for a minimal input perturbation that flips the label and then fine‑tunes on the resulting corrected pair~\cite{Wang2019}. Methods like Implicit Backdoor Adversarial Unlearning (\emph{I‑BAU})~\cite{Zeng2022IBAU}, Anti-Backdoor Learning (\emph{ABL})~\cite{li2021abl}, and Neural attention distillation \emph{(NAD)}~\cite{Li2021NAD} extend this approach to broader trigger types, while more recently, neural‑collapse cleansing restores a backdoored network by realigning its feature geometry with that of a clean reference model~\cite{Gu2024}.
\noindent
While \emph{NAD} assumes access to a clean teacher, \emph{PEFTGuard}~\cite{Sun2025PEFTGuard} facilitates backdoor localization for parameter-efficient settings. It automatically inspects LoRA and adapter modules for potential backdoor logic and provides saliency-style visualizations of their effect on model behavior. \emph{PEFTGuard} combines lightweight structural analysis with feature attribution to identify which modules disproportionately influence attacker-desired outputs. Although designed for PEFT scenarios, its logic-based analysis generalizes well to architectural backdoor repair tasks involving sparse or modular subgraphs.
\noindent
Adversarial unlearning is effective primarily for backdoors that actively influence model parameters or decision boundaries. Architectural backdoors, especially those rarely activated, present limited gradient signals during fine-tuning, rendering unlearning incomplete. Composite or multi-condition triggers require exposing every trigger condition, further complicating remediation efforts. More fundamentally, cryptographic backdoors compiled into the model architecture, as shown by Draguns \textit{et al.}~\cite{Draguns2025Unelicitable}, may resist all current unlearning methods. Their encrypted trigger-payload circuits cannot be reliably activated even through latent adversarial training (LAT), which is widely viewed as the state of the art in backdoor elicitation. This suggests that adversarial fine-tuning cannot mitigate backdoors that are non-continuously differentiable or intentionally obfuscated at the computational-graph level. These limitations motivate complementary, lightweight defenses such as pruning, attention distillation (§\ref{sec:mitigation-distillation}), and runtime monitoring (§\ref{sec:mitigation-runtime}).
\noindent
Recent work on securing transfer learning pipelines introduces proactive filtering to exclude compromised components before they are learned. \emph{T-Core Bootstrapping}~\cite{Zhang2025SecureTransfer} identifies trustworthy neurons and data instances early in fine-tuning, mitigating both architectural and parameter-based backdoors in pre-trained encoders. By constraining the transfer learning process to “core-safe” features, \emph{T-Core} reduces the risk that a dormant backdoor is inherited from a contaminated upstream model or dataset. This strategy is particularly valuable when retraining on sensitive downstream tasks where subtle misbehavior could evade manual inspection.

\subsection{Attention Distillation and Model Surgery}
\label{sec:mitigation-distillation}

\emph{NAD} aligns a potentially backdoored student model to a clean teacher by matching intermediate attention maps~\cite{Li2021NAD}. Li~\textit{et al.} demonstrated that \emph{NAD} significantly recovers clean accuracy. However, \emph{NAD}’s effectiveness depends critically on having access to a trusted clean teacher model of similar architecture, a prerequisite often unavailable in practice. Further, architectural backdoors that span multiple distributed branches may evade partial realignment or isolated surgical interventions, necessitating more extensive model surgery or combined mitigation strategies.
\noindent
If the malicious logic sits in identifiable heads or layers, simply
disabling those modules and fine‑tuning can work
\cite{pang2022,Langford2025}. Multi-branch attacks often do not localize neatly into one identifiable head or layer, requiring more extensive surgery or combination with other mitigation strategies to avoid significant accuracy loss. Combined with subgraph excision (§\ref{sec:mitigation-pruning}), these attention‑centric methods form the second line of defense before costly full re‑training or supply‑chain rebuilds.

\begin{table}[htbp]
  \caption{\textbf{Attention-based repair: pros and cons}}
  \label{tab:nad-pros-cons}
  \centering
  \footnotesize                      % matches the other tables
  \begin{tabularx}{\linewidth}{P{2.7cm} Y Y}
    \toprule
     & \textbf{Strengths} & \textbf{Caveats}\\
    \midrule
    \emph{NAD} &
      Systematically realigns internal attention; no full retrain if teacher available &
      Needs a clean teacher of similar architecture \\ \midrule
    Layer/Head removal &
      Fast; surgically targets known gate &
      Accuracy drop if gate overlaps benign function; ineffective on interwoven logic \\
    \bottomrule
  \end{tabularx}
\end{table}

% =============================================================

\subsection{Runtime Monitoring and Canary Testing}
\label{sec:mitigation-runtime}

Stealthy architectural backdoors may activate only on rare inputs, so \emph{in-production} monitoring must complement offline defences.

\paragraph{Entropy / anomaly monitors \& canaries.}
Integrate STRIP-style entropy checks~\cite{gao2019strip} and inject crafted canaries that mimic likely triggers~\cite{Clifford2024}. Each flags a live sub-graph, yet composite or distributed triggers~\cite{xia2024,Langford2025} can evade both, so runtime monitoring should be viewed as a layer, not a fix.

\paragraph{Input randomisation.}
Pixel or gate noise can break exact-match triggers~\cite{zeng2022,pang2022}; robust backdoors survive, stressing the need for multi-pronged runtime guards.

\paragraph{Domain-specific runtime shields.}
\emph{GraphProt}~\cite{Yang2024GraphProt} defends black-box graph classifiers by (i) clustering each input graph, (ii) sampling purified sub-graphs, and (iii) ensembling their predictions, cutting ASR by up to 90\% on six datasets with around 1–2\% accuracy loss.  Because it needs no retraining or weight access, \emph{GraphProt} illustrates practical, domain-aware runtime shielding.

\paragraph{LLM-assisted filtering for recommender systems.}
\emph{P-Scanner}~\cite{Ning2025PScanner} counters \textsc{BadRec}, a 1\%-poison backdoor that forces any tokenized item to be recommended, by fine-tuning a large language model to spot semantic anomalies.  It removes > 90 \% of poisoned samples across three real-world datasets, restoring accuracy with negligible overhead.

Runtime monitoring flags, but does not remove, a backdoor; it is most effective alongside the offline mitigations of §§\ref{sec:mitigation-pruning}–\ref{sec:mitigation-distillation}.  Structural threats already permeate diffusion models too~\cite{Chou2023BadDiffusion}, underscoring the need for broad, layered runtime defenses.

\begin{table}[htbp]
  \caption{\textbf{Runtime checks: strengths and caveats}}
  \label{tab:runtime-pros-cons}
  \centering
  \footnotesize                 
  \begin{tabularx}{\linewidth}{P{2.4cm} Y Y}
    \toprule
    & \textbf{Strengths} & \textbf{Limitations} \\
    \midrule
    Entropy/anomaly &
      Low overhead; continuous monitoring &
      False negatives if outputs stay “plausible” \\ \midrule
    Canary testing &
      Direct confirmation of malicious path &
      Requires guessing the trigger; composite keys may evade \\ \midrule
    Randomization &
      Cheap perturbation defense &
      Advanced backdoors can be perturbation-robust \\
    \bottomrule
  \end{tabularx}
\end{table}

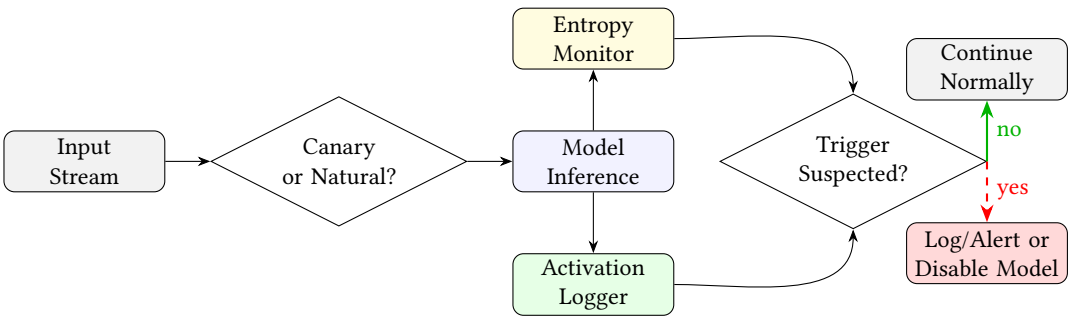
\begin{figure}[htbp]
\centering
\begin{tikzpicture}[
    font=\small,
    node distance=0.8cm and 0.6cm,
    >=Stealth,
    diamondbox/.style={
        draw,
        diamond,
        aspect=2,
        text width=2cm,
        align=center,
        inner sep=1pt
    },
    rectbox/.style={
        draw,
        rectangle,
        rounded corners,
        align=center,
        minimum height=0.8cm,
        text width=2cm,
        inner sep=2pt
    }
]

% 1) Input at the far left
\node[rectbox, fill=gray!10] (input) {Input\\Stream};

% 2) Diamond: Canary or Natural?
\node[diamondbox, right=of input] (split) {Canary\\or Natural?};

% 3) Model Inference
\node[rectbox, fill=blue!5, right=of split] (model) {Model\\Inference};

% 4) Entropy/Confidence Monitor (above model)
\node[rectbox, fill=yellow!15, above=of model] (entropy) {Entropy\\Monitor};

% 5) Activation Logger (below model)
\node[rectbox, fill=green!10, below=of model] (activations) {Activation\\Logger};

% 6) Decision: Trigger Suspected?
\node[diamondbox, right=of model] (decision) {Trigger\\Suspected?};

% 7) Alert/Disable if yes
\node[rectbox, fill=red!15, below=of decision.east] (flag) {Log/Alert or\\Disable Model};

% 8) Continue if no
\node[rectbox, fill=gray!10, above=of decision.east] (safe) {Continue\\Normally};

% - Arrows -
\draw[->] (input) -- (split);
\draw[->] (split) -- (model);

%\draw[->] (model.east) -- (decision.west);

% Entropy above model => to decision
\draw[->] (model.north) -- (entropy.south);
\draw[->] (entropy.east) to[out=0,in=90] (decision.north);

% Activations below model => to decision
\draw[->] (model.south) -- (activations.north);
\draw[->] (activations.east) to[out=0,in=-90] (decision.south);

% Decision outcome (colour-only changes)
\draw[->, red, dashed, thick]   (decision.east) -- node[midway,right]{\small yes} (flag.north);
\draw[->, green!70!black, thick](decision.east) -- node[midway,right]{\small no}  (safe.south);

\end{tikzpicture}
\caption{\label{fig:runtime_detection_flow}
\textbf{Runtime Backdoor Detection Flow (rotated).}
Even after offline mitigation, defenders can feed canary inputs or monitor normal traffic for suspicious triggers. If detected, the model can be logged or sandboxed; otherwise, 
inference continues normally.}
\Description{Vertical flowchart showing an input stream passing through a canary/normal branch, an entropy monitor, and an activation logger if a trigger is suspected, the system logs or disables the model.}
\end{figure}

 %=============================================================
\subsection{Supply‑Chain Assurance and Trusted Compilation}
\label{sec:mitigation-supplychain}

Even after local repair, a tainted build pipeline can surreptitiously re-insert malicious logic at export time.  
Recent incident reports from Protect AI (\texttt{PAIT-ONNX-200} and \texttt{PAIT-TF-200}) uncovered ONNX and TensorFlow SavedModel artefacts whose hidden trigger branches were added only during serialization \cite{protectai2024PAIT}.  
Zhu \textit{et al.} further show that TensorFlow’s export APIs can embed shell-executing ops into the graph, transforming an ordinary model into a malware container without altering predictions \cite{Zhu2024TFMalware}. These findings underscore that robust assurance must span the entire pipeline: \emph{source → compiler IR → runtime → hardware}. Deterministic builds enable IR differencing, and recent static–taint methods such as Batch Isolation Checker attach a short, machine-checkable proof to each lowering pass~\cite{Kuchler2025BatchSteal}.
\noindent
See §\ref{sec:future} for a detailed checklist on reproducible builds and IR-level provenance.

% \paragraph{Reproducible builds and IR diffs.} Signed deterministic compilation and reproducible build systems ensure each step hashes to a cryptographic reference, allowing IR-level comparisons (e.g., ONNX diff) against a trusted baseline. Practitioners should leverage tools such as Bazel’s reproducible builds~\cite{bazel2024reproducible, zheng2024bazelci} or container-based solutions like ReproZip~\cite{chirigati2016reprozip}, and adopt best practices from ML provenance frameworks (e.g., Model Cards, FactSheets, or NIST’s AI risk management guidelines). Any IR discrepancy could indicate potential compiler-level tampering, underscoring the necessity of verifying builds at each stage of the pipeline.

\paragraph{Malicious AutoML pipelines.} Attacks such as \emph{DarkNAS} embed backdoors during architecture search when the reward signal is under adversarial control \cite{pang2022,pang2023dark}.  Supply‑chain checks must therefore verify AutoML artifacts as well as compiler outputs.

\paragraph{TensorFlow logic-bomb exploits.}  
Zhu \textit{et al.} demonstrate that only a handful of TensorFlow API calls, inserted during export, can plant executable payloads inside a SavedModel graph \cite{Zhu2024TFMalware}.  
The malicious ops fire at deserialization time, executing arbitrary commands while leaving model accuracy intact.  
Because the backdoor lives in the standard SavedModel format, it evades conventional weight checks and illustrates why supply-chain audits must include static operator allow-lists and sandboxed loading for untrusted artefacts.

\paragraph{Hardware Trojan risk.} Compromised accelerators can manifest a backdoor at run‑time without changing weights \cite{Sengupta2025HardwareTrojanML}. End‑to‑end assurance echoes SoC security practice \cite{tehranipoor2010survey}. A key takeaway from the Logic backdoor proof-of-concept~\cite{hiddenlayer2024} is that stealth subgraphs added post-training may go unnoticed in source code and only be revealed through IR differencing. This highlights compiler-time insertion as a potent and easily overlooked infiltration vector.

\paragraph{Replicated Execution for Outsourced Training}
%\label{sec:replicated-execution}
Replicated execution across multiple non-colluding cloud providers~\cite{jia2025outsourced} can verify outsourced training jobs by cross-comparing intermediate checkpoints for anomalies. Although redundant execution increases costs, for critical models, training smaller subset models across diverse providers can effectively identify suspicious insertions. Federated learning frameworks, which already employ partial model averaging and integrity checks, provide a relevant precedent illustrating this approach's feasibility in practice.

% --
%\subsection{Comparative Overview of Mitigations}
%\label{sec:discussion-mitigation}
\subsection{Concluding Summary}
\label{sec:mitigation-conclusion}

\begin{table}[htbp]
  \caption{\textbf{Comparison of Mitigation Methods for Architectural Backdoors.}
    Each approach tackles a different slice of backdoor logic; multi-branch or
    compiler trojans often need a mix of these methods.}
  \label{tab:mitigation_comparison}
  \centering
  \footnotesize
  \begin{tabular*}{\columnwidth}{@{\extracolsep{\fill}}
          P{2.2cm} P{5.1cm} P{5.1cm}}
    \toprule
    \textbf{Method} & \textbf{Key Advantages} & \textbf{Main Limitations} \\ \midrule
    \textbf{Subgraph Removal} &
      Physically excises identified malicious routes; low data cost; effective
      if the backdoor path is separate. &
      Fails for integrated triggers woven into normal blocks; may hurt accuracy
      if removal is too extensive. \\ \midrule
    \textbf{Adversarial Unlearning \& Re-Training} &
      Disassociates known triggers from attacker outputs; flexible across
      backdoor types; I-BAU cuts overhead. &
      Must activate every trigger to “unlearn’’ it; multi-trigger logic can
      persist; costly for large models. \\ \midrule
    \textbf{Attention Distillation \& Surgery} &
      Re-aligns suspicious layers/heads; partial fine-tune preserves benign
      performance if a clean teacher exists. &
      Distributed or multi-branch trojans may evade partial realignment; need a
      precise mapping of malicious parts. \\ \midrule
    \textbf{Runtime Monitoring \& Canary Testing} &
      Watches models in deployment; can catch unforeseen triggers; little extra
      training. &
      Does not remove the trojan; stealthy gating can mimic normal outputs and
      slip past anomaly checks. \\ \midrule
    \textbf{Supply-Chain Assurance} &
      Blocks re-injection; guarantees final artefact matches a trusted build;
      covers AutoML or compiler trojans. &
      Adds organisational overhead; needs signed/reproducible builds; hardware
      Trojans remain a separate threat. \\ \bottomrule
  \end{tabular*}
  \Description[Mitigation comparison table]{Key advantages and limitations of five mitigation methods.}
\end{table}

\begin{figure}[htbp]
\centering
\resizebox{0.9\textwidth}{!}{%
\begin{tikzpicture}[
  node distance=1.1cm and 2.2cm,
  safe/.style={draw, rectangle, rounded corners, fill=green!20, 
               minimum width=4cm, minimum height=0.9cm, 
               font=\small, align=center},
  action/.style={draw, rectangle, rounded corners, fill=red!15, 
                 minimum width=4cm, minimum height=0.9cm, 
                 font=\small, align=center},
  decision/.style={draw, diamond, aspect=2.2, fill=yellow!20, 
                   minimum width=3.5cm, minimum height=1.0cm, 
                   font=\small, align=center, thick},
  arrow/.style={->, >=stealth, thick},
  font=\small
]

% Top-level decision
\node[decision] (suspicion) {Suspicious architecture?};

% Second level
\node[decision, below left=1.4cm and 2.2cm of suspicion] (static) 
  {Static check:\\ unusual sub-ops?};
\node[decision, below right=1.4cm and 2.2cm of suspicion] (behavior) 
  {Trigger-like behavior\\ on test data?};

% Third level
\node[action, below=0.4cm of static] (remove_branch) 
  {Remove/disable\\malicious subgraphs};
\node[action, below=1.4cm of behavior] (synth_trigger) 
  {Trigger synthesis\\(Neural Cleanse, etc.)};

% Fourth level
\node[decision, below=1.6cm of remove_branch] (impact) 
  {Clean accuracy unchanged?};

% Final outcomes
\node[safe, below left=1.2cm and 1.5cm of impact] (accept_clean) 
  {Mitigation successful};
\node[action, below right=1.2cm and 1.5cm of impact] (retrain_variant) 
  {Retrain \& incorporate\\ discovered triggers};

% Final branch
\node[action, below=1.8cm of synth_trigger] (fineprune) 
  {Apply adversarial\\ unlearning or surgery};

% Arrows
\draw[arrow] (suspicion) -- (static);
\draw[arrow] (suspicion) -- (behavior);
\draw[arrow] (static) -- (remove_branch);
\draw[arrow] (behavior) -- (synth_trigger);
\draw[arrow] (remove_branch) -- (impact);
\draw[arrow, green!70!black, thick] (impact) -- (accept_clean);      % success path
\draw[arrow, red, dashed, thick]    (impact) -- (retrain_variant);   % further-action path

\draw[arrow] (synth_trigger) -- (fineprune);

\end{tikzpicture}%
} % End resizebox
\caption{\label{fig:mitigation_decision_tree}
\textbf{Decision flow for mitigating architectural backdoors.} 
Starting from suspected infiltration, defenders apply static or dynamic checks to pinpoint malicious subgraphs or triggers. 
They then perform subgraph removal, adversarial unlearning, or attention surgery. 
If clean accuracy is significantly impacted, more extensive re-training follows. 
Finally, ensuring a trusted supply chain (§\ref{sec:mitigation-supplychain}) prevents backdoor re-insertion.}
\Description{Yes/No decision tree that starts with “Suspicious architecture?” and routes through static checks, trigger synthesis, subgraph removal, unlearning, or full retraining, ending at “Mitigation successful.”}
\end{figure}
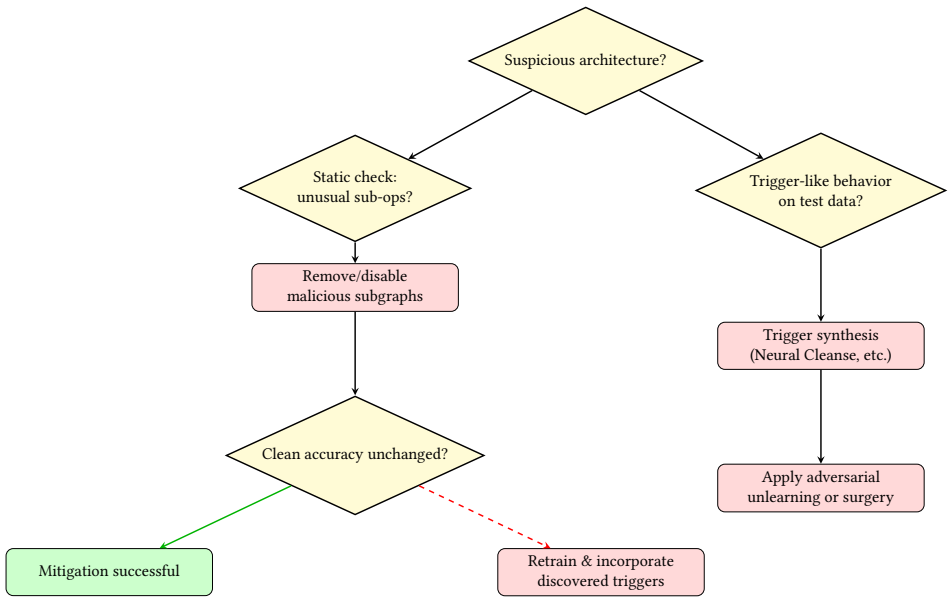

%\subsection{Concluding Summary and Future Directions}
%\label{sec:mitigation-conclusion}
\noindent
Architectural backdoors reside in \emph{topology}, so defenders must layer remedies: subgraph excision, adversarial unlearning, attention
surgery, runtime checks, and critically-supply-chain signing.
Table~\ref{tab:mitigation_comparison} summarizes how each technique addresses architectural backdoor; Figure~\ref{fig:mitigation_decision_tree} gives one possible decision flow.

% TODO: smoothe out this paragraph
\paragraph{Open problems.}
(1) Fully integrated multi‑trigger attacks still defeat current tools. 
(2) Scalable formal repair for billion‑parameter graphs remains elusive (§\ref{sec:formal-verification}).  
(3) Pipeline re‑introduction demands continuous verification, not one‑off cleaning.  
(4) Benchmark gaps slow progress; few suites model compiler‑time insertion or AutoML backdoors.
\noindent
Continued cross‑layer research is needed for durable defense: this must span graph analysis, automated repair, secure AutoML, reproducible builds, and hardware attestation. Addressing these open problems will require sustained interdisciplinary collaboration among researchers, industry practitioners, and regulatory bodies, ensuring that advancements in graph-level verification, runtime defenses, secure compilation, and hardware integrity checks are tightly integrated into a robust, end-to-end defense ecosystem against architectural backdoors.

% =============================================================

\section{Benchmarks, Datasets, and Empirical Evaluations}
\label{sec:benchmarks}

Having surveyed detection (\ref{sec:detection}) and mitigation (\ref{sec:mitigation}), we now examine how well those defenses hold up in practice. Although many benchmark suites exist for data‑ or weight‑centric backdoors, few systematically address structural or compiler‑level attacks. This section reviews current resources, highlights gaps, and suggests directions for multi‑branch and supply‑chain‑aware evaluations.

% Possible new structure:
%  - Current State of Backdoor Benchmarking: "Context" (6.2) + "Major Benchmarks" (6.3) + Case Study (6.7)
%  - Mitigation success metrics for architechtural backdoors: "Quanitative coverage" (6.4) + "Performance Under Retraining" (6.1) + Distinctions" (6.5) + "repair benchmarks" (6.12)
%   - Emerging challenges: complex architectures like LLMs (6.6), formal verification (6.8), repositories (6.9), adaptive attacks (6.10)
%   - Future suggestions: "proposed solutions" (6.11) + "summary" (6.13)

%------------------------------
\subsection{Current State of Backdoor Benchmarking}
\label{sec:benchmark-status}

Standard backdoor metrics (detection TPR/FPR, ASR reduction) still apply, but evaluations must treat the trigger as a latent sub-graph or operator sequence rather than a visible pixel patch. \emph{TrojAI} Round 15 was the first public benchmark to distribute ONNX binaries that contain hidden branches requiring structural detection or neutralization, all without relying on poisoned weights~\cite{trojai-round15}. Nevertheless, coverage is still incomplete: pure compiler-time inserts~\cite{Clifford2024}, multi-branch triggers, and post-fine-tune reactivation remain largely untested.
\noindent
Protect AI’s PAIT scans flag real ONNX/SavedModel artefacts with hidden branches (e.g.\ \texttt{PAIT-ONNX-20}, \texttt{PAIT-TF-200}) and publish full metadata~\cite{ProtectAI2025SixMonthReport}.  Incorporating a “PAIT track’’ in future rounds (\emph{TrojAI} R14, \emph{BackdoorBench}-X) would inject live, diverse compiler-time trojans and tighten realism.
\noindent
\emph{BackdoorBench} adds only a few gating-op tests; \emph{TrojAI} and the NeurIPS’20 challenge target data/weight triggers~\cite{NeurIPSTrojDet2020}.  Ad-hoc academic sets rarely include compiler or multi-branch inserts~\cite{Bober-Irizar2023,Pang2022AutoMLRisk}, nor do they assess fine-tuning persistence or supply-chain infiltration.  Table~\ref{tab:benchmarks} summarizes present coverage.

\begin{table}[htbp]
  \caption{Architectural coverage of today’s backdoor benchmarks}
  \label{tab:benchmarks}
  \footnotesize
  \centering
  \begin{tabular*}{\linewidth}{@{\extracolsep{\fill}}
          P{3.2cm} P{4.3cm} P{3.2cm}}
    \toprule
    \textbf{Benchmark} & \textbf{Main focus} & \textbf{Struct./Compiler coverage} \\
    \midrule
    TrojAI (IARPA) & Data‐poison, weight triggers & Minimal; no compiler path \\ \midrule
    BackdoorBench & Data/weight + a few gating ops & No multi-branch or IR sabotage \\ \midrule
    NeurIPS TD Challenge & Competition on parameter triggers & None (structural absent) \\ \midrule
    Academic sets & Small crafted examples & Sporadic, ad-hoc structures \\
    \bottomrule
  \end{tabular*}
\end{table}

\noindent
Due to the lack of coverage for architectural backdoors in most backdoor benchmarks, many architectural-backdoor studies rely on ad-hoc model sets. Bober-Irizar \emph{et al.}~\cite{Bober-Irizar2023} used about 20 AlexNet variants with a checkerboard trigger, while Pang \emph{et al.}~\cite{pang2022} manipulated a NAS pipeline for CIFAR. Both illustrate key ideas but lack scale or diversity comparable to data/weight-centric benchmarks. These examples underscore an urgent need for more extensive, standardized evaluations tailored to structural infiltration.

%------------------------------
\subsection{Measuring Success in Architectural Backdoor Mitigation}
\label{sec:benchmark-backdoor-mitigation-metrics}
%  - Mitigation success metrics for architechtural backdoors: "Quanitative coverage" (6.4) + "Performance Under Retraining" (6.1) + Distinctions" (6.5) + "repair benchmarks" (6.12)

Mainstream benchmarks typically report true/false positive rates, ASR reduction, and resource overhead. Although these metrics work for architectural backdoors, they do not capture unique complexities like multi-branch triggers or domain-shift reactivation. 
A hallmark of architectural backdoors is their survivability, they often persist after full re‑training.
For instance, backdoor studies based on the \emph{CIFAR-10} dataset demonstrated that weight-based backdoors can be “forgotten” by retraining, but model architecture backdoors (MABs)~\cite{Bober-Irizar2023} and operator-based~\cite{Langford2025}) retained attack success even after full retraining.
Similarly, Gu \emph{et al.}~\cite{Gu2017badnets} showed that a trigger in a traffic-sign classifier persisted after fine-tuning on a new task, an aspect seldom tested in standard frameworks.
Attack success for models with architectural backdoors remains high in these studies because the malicious route is encoded in the model’s graph, not just in weights.

% distinctions + repair benchmarks
\noindent
Beyond detection, effective mitigation needs benchmarks with known malicious substructures, enabling measurements of “repair completeness” and performance overhead. No mainstream resource yet includes detailed annotations for structural backdoors. 
Current benchmarks rarely address multi-branch gating, compiler-level infiltration~\cite{Clifford2024}, or AutoML-based Trojan designs~\cite{pang2022,pang2023dark}. Classification tasks (e.g., \emph{CIFAR-10}, \emph{ImageNet}) dominate, overlooking more complex applications like segmentation or high-performance computing (HPC)-scale pipelines. Even fewer consider “post-training insertion” scenarios where malicious logic is introduced into ONNX files after the model is published.
Bridging legacy data/weight sets with newly introduced structural infiltration could help the community assess repair strategies more rigorously.

\subsection{Proposed Benchmarking Solutions}
\label{sec:benchmark-solutions}
%   - Future suggestions: "proposed solutions" (6.11) + "summary" (6.13)
To move from ad-hoc structural demos to rigorous evaluation, the community needs purpose-built benchmarks addressing multi-branch gating, compiler-level infiltration, and supply-chain reinfection. Table~\ref{tab:benchmark-roadmap} outlines four potential directions:

\begin{itemize}
    \item \textbf{Extend existing suites (\emph{TrojAI, BackdoorBench}):} Incorporate multi‑branch, compiler-time, and large-scale LLM tasks using established scoring pipelines.
    \item \textbf{Repair benchmarks:} Provide ground-truth subgraph labels, enabling fair comparison of excision, unlearning, or distillation methods.
    \item \textbf{Arms-race tracks:} Host periodic competitions that evolve attacks and defenses, preventing over-fitting to static scenarios.
    \item \textbf{Industry \& hub collaboration:} Aggregate suspicious uploads from real-world repositories. Overcome legal/licensing barriers to build a comprehensive, publicly accessible dataset.
\end{itemize}

\begin{table}[htbp]
  \caption{Roadmap for next-generation benchmarks on architectural backdoors.}
  \label{tab:benchmark-roadmap}
  \centering
  \footnotesize
  \begin{tabularx}{\linewidth}{P{2.6cm} Y Y}
    \toprule
    \textbf{Initiative} & \textbf{Scope \& Expected Gain} & \textbf{Key Obstacles}\\
    \midrule
    \textbf{Expand legacy suites} (TrojAI, BackdoorBench) &
      Add multi-branch models, compiler inserts, LLM or multi-modal tasks &
      New generation scripts; licensing of non-open IRs.\\
    \midrule
    \textbf{Repair benchmarks} &
      Ground-truth subgraph masks for subgraph-excision methods &
      Annotating malicious ops at scale; preserving benign performance.\\
    \midrule
    \textbf{Arms-race tracks} &
      Annual events that evolve attacks/defenses &
      Organizer bandwidth; dynamic rules for success.\\
    \midrule
    \textbf{Industry \& hub collaboration} &
      Live scanning of user-submitted models &
      Privacy/legal issues; inconsistent licensing; incentives for submission.\\
    \bottomrule
  \end{tabularx}
\end{table}

\noindent
Extensive expansion of backdoor benchmarks is needed to address architectural backdoors. Table~\ref{tab:benchmarks} shows that recognized programs like \emph{TrojAI} and \emph{BackdoorBench} devote minimal attention to multi-branch or compiler-time triggers. Although a few works address LLM or multi-modal vulnerabilities, none provide large-scale, community-driven benchmarks. An adaptive approach that updates structural infiltration scenarios regularly, coupled with academic-industry collaboration on real repository scans, would keep defenders aligned with evolving adversarial techniques. We thus call for dedicated tasks, or expansions within popular frameworks, focusing on multi-branch gating, compiler-level insertions, and domain-shift reactivation, ensuring defenses are tested against the full spectrum of architectural backdoors.

\section{Open Challenges and Future Directions}
\label{sec:future}

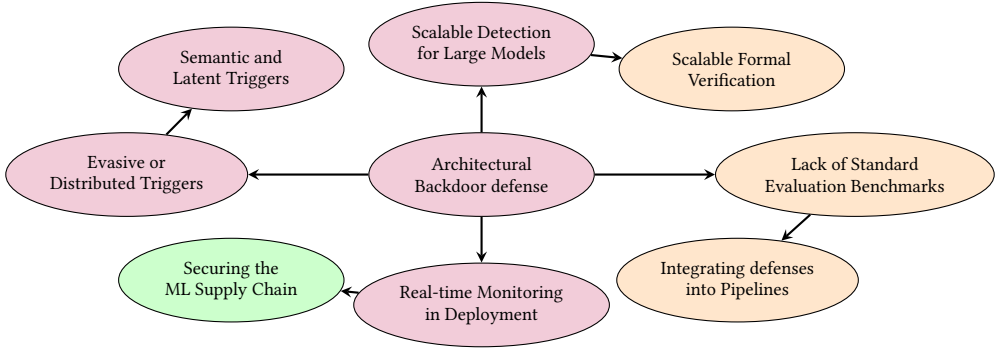
\begin{figure}[htbp]
\centering
\begin{tikzpicture}[
  tech/.style   ={draw, ellipse, fill=purple!20,  % ← purple to match caption
                  minimum width=3.0cm, minimum height=1.1cm,
                  font=\scriptsize, align=center},
  method/.style ={draw, ellipse, fill=orange!20,
                  minimum width=3.0cm, minimum height=1.1cm,
                  font=\scriptsize, align=center},
  ops/.style    ={draw, ellipse, fill=green!20,
                  minimum width=3.0cm, minimum height=1.1cm,
                  font=\scriptsize, align=center},
  link/.style={->, >=stealth, thick},
  node distance=0.6cm and 1.6cm,
  font=\scriptsize
]

% Central node
\node[tech] (centre) {Architectural\\Backdoor defense};

% Inner ring
\node[tech, above=of centre] (scaling) {Scalable Detection\\for Large Models};
\node[tech, below=of centre] (runtime) {Real-time Monitoring\\in Deployment};
\node[tech, left=of centre]  (evasion) {Evasive or\\Distributed Triggers};
\node[method, right=of centre] (benchmark) {Lack of Standard\\Evaluation Benchmarks};

% Outer ring
\node[ops, below left=0.6cm and 1.2cm of centre] (supply) {Securing the\\ML Supply Chain};
\node[method, above right=0.6cm and 1.2cm of centre] (formality) {Scalable Formal\\Verification};
\node[tech, above left=0.6cm and 1.2cm of centre] (semantics) {Semantic and\\Latent Triggers};
\node[method, below right=0.6cm and 1.2cm of centre] (integration) {Integrating defenses\\into Pipelines};

% Links
\draw[link] (centre) -- (scaling);
\draw[link] (centre) -- (runtime);
\draw[link] (centre) -- (evasion);
\draw[link] (centre) -- (benchmark);

\draw[link] (evasion)  -- (semantics);
\draw[link] (scaling)  -- (formality);
\draw[link] (benchmark)-- (integration);
\draw[link] (runtime)  -- (supply);

\end{tikzpicture}
\caption{\textbf{Conceptual map of open challenges in defending against architectural backdoors.}}
\label{fig:challenges_map}
\Description{Radial concept map with purple, orange, and green nodes (e.g.\ “Scalable Detection,” “Formal Verification,” “Supply-chain Security”) connected by arrows to show inter-dependencies among research gaps.}
\end{figure}

\noindent
Architectural backdoors present threats that conventional data-poisoning or parameter-centric defenses rarely address, especially when embedded at the compiler or hardware level. 
The preceding sections (§\ref{sec:detection}--§\ref{sec:benchmarks}) have analyzed subgraph modifications, AutoML vulnerabilities, and the limited scope of current benchmarks. Despite these advances, several obstacles remain before comprehensive security against architectural backdoors can be achieved, particularly at scale. This section highlights the open challenges, then proposes directions for large-scale verification frameworks, supply-chain governance, and adaptive benchmarking, ultimately aiming for robust, end-to-end solutions that anticipate emerging infiltration tactics. Figure~\ref{fig:challenges_map} summarizes how the major research and operational gaps inter‑relate; the subsections that follow unpack each edge of the diagram in depth.

% Revised Section 7.1 – Overview of Core Challenges (merged with former §6.3)
% -----------------------------------------------------------------------
\noindent
\textbf{Summary of open challenges and future directions:}
\begin{enumerate}
    \item \textbf{Scaling Formal Methods and Large-Scale Verification:}  Satisfiability Modulo Theories (SMT)-based and abstract-interpretation verifiers achieve promising guarantees on feed-forward models of tens of millions of parameters ($10^7$), but attempts to push them to Transformer-scale routinely time-out or blow up in memory~\cite{Pham2022CAV}.  \emph{How can compositional, approximate, or statistical verification frameworks certify safety properties of trillion-parameter or HPC-scale architectures without prohibitive cost?}

    \item \textbf{Addressing Multi-Path, Distributed, or Obfuscated Triggers:} Hidden backdoor logic is increasingly split across multiple attention heads, residual branches, or even modalities; triggers may require specific chain-of-thought (CoT), token sequences, or coordinated text–image pairs.  Recent studies into distributed triggers~\cite{Langford2025}, frequency-obfuscated vision backdoors (\emph{LADDER}~\cite{liu2025ladder}), CoT hijacks such as \emph{BadChain}~\cite{xiang2024badchain}, \emph{DarkMind}~\cite{DarkMind2025}, \emph{ShadowCoT}~\cite{Zhao2025ShadowCoT}, \emph{TransTroj} fine-tuning persistence~\cite{Wang2025TransTroj}, and the open-source \emph{BackdoorLLM} pipeline~\cite{Li2024BackdoorLLM}, achieve >90\% attack success while bypassing current detectors, and no public benchmark yet exercises these cross-modal gates~\cite{Zhao2025LLMBackdoors}. \emph{What graph-analytic or constraint-solving techniques can reason about distributed activation patterns and reliably surface stealthy multi-branch trigger combinations?}

    \item \textbf{NAS and AutoML Vulnerabilities:} NAS and other AutoML pipelines can inadvertently embed, or deliberately plant, malicious sub-structures that optimize for a poisoned reward signal. \emph{How can integrity checks on reward functions, search spaces, and validation loops guarantee that generated architectures are provably free of covert backdoor logic?}

    \item \textbf{Compiler/Hardware Synergy and Post-Training Backdoor Re-Introduction:} Malicious compilers or hardware Trojans can resurrect excised subgraphs after software-level cleansing via post-training reinsertion (e.g., \emph{Shadow Logic}~\cite{hiddenlayer2024}) or IR manipulation, recreating backdoors at deploy time~\cite{Clifford2024}. \emph{Can we design end-to-end trusted toolchains that preserve a verifiable equivalence between audited source graphs and the binaries and silicon ultimately deployed?}

    \item \textbf{Adaptive Benchmarks and Continuous Updating:}  Static challenge suites quickly become outdated as adversaries evolve.  Preliminary hub-scanning tools such as \emph{CLIBE} (deteCting NLP dynamIc Backdoor TransformEr)~\cite{zeng2024clibe} remain proprietary, and no open governing body curates an evolving benchmark.  \emph{What incentive structures would support a living benchmark,  including real-world corpus tracks and LLM-orchestrated defender baselines~\cite{Castro2025LLMDefenders}, that stay synchronized with emerging attack strategies?}

    \item \textbf{Specialized Architectures: Spiking Neural Networks (SNNs) and Visual State-Space Models (VSSMs):}  Non-canonical computing graphs diverge from CNN/Transformer assumptions, creating blind spots for existing defenses.  \emph{Which abstraction layers or surrogate models will let us reason about temporally-coded or state-space activations without sacrificing threat coverage?}

    \item \textbf{Supply-Chain Governance, Policy, and Multi-Sector Collaboration:}  Public hubs and proprietary exchanges lack harmonized auditing standards, giving attackers ample surface for injection.  Successful large-scale defense will likely require joint policy, open-source tooling, and certification programs akin to software bill of material (SBOM) initiatives.  \emph{What regulatory, policy, and technical levers can align incentives across academia, industry, and government to secure the global model supply chain?}
\end{enumerate}

\noindent
Recent empirical work on GPT-style models by Miah and Bi~\cite{Miah2024ArchLLM}, together with the comprehensive LLM survey by Zhao \emph{et al.}~\cite{Zhao2025LLMBackdoors}, confirms that architectural backdoors have already migrated to billion-parameter language models.

    % ---------------------------------------------------------------
    \subsection{Scaling Formal Methods and Large‑Scale Verification}
    \label{sec:future-verif}
    
    \textbf{Problem statement:}  Formal, symbolic, and abstract‑interpretation–based verifiers can certify small neural networks against backdoor behavior under bounded perturbations\cite{Katz2017Reluplex,gehr2018ai2,eran2020tool,singh2019deeppoly}.  Beyond roughly $10^7$ parameters they either time‑out or excessively over‑approximate, leaving modern CNNs, vision–language transformers, and all contemporary LLMs effectively unverifiable.  Compiler‑time backdoor re‑insertion, demonstrated by \emph{ImpNet}/\emph{Shadow Logic} attacks\cite{Clifford2024}, further undercuts proof soundness: a source graph proven clean can still be corrupted during lowering to inference binaries.
    
    \noindent
    \textbf{Why current defenses fall short}
    \begin{itemize}[leftmargin=1.5em,itemsep=2pt]
      \item \emph{State‑space explosion.}  Parameter counts scale super‑linearly with depth and width; existing SMT or Mixed Integer Linear Programming (MILP) encodings become intractable for trillion‑parameter models or multi‑branch architectures with distributed gating triggers.
      \item \emph{Compiler/Hardware gap.}  Proofs apply to the abstract graph prior to optimization passes; malicious compilers or rogue hardware IP can violate the verified property post‑compile, creating an unmonitored attack surface.
      \item \emph{Lack of incremental benchmarks.} Public challenge suites rarely include models $>100$M parameters, so progress on scaling techniques is neither tracked nor rewarded.
      
    \end{itemize}
    \noindent
    \textbf{Promising research directions}
    \begin{itemize}
    
    \item \textit{Compositional proofs.} Decompose networks into verifiable blocks (layers, residual paths) and stitch guarantees via assume–guarantee reasoning \cite{Pasareanu2018Compositional} or category-theoretic lenses \cite{Fong2019BackpropFunctor}. Partial coverage can still block many distributed trigger designs.

      \item \textit{HPC‑assisted verification.}  Cluster‑scale symbolic execution frameworks, coupled with aggressive post‑pruning (§\ref{sec:mitigation-pruning}), could deliver tractable yet conservative proofs for very large models.
      \item \textit{Runtime attestation.}  Lightweight on‑device monitors can hash or attest critical gating operations during inference, providing continuous assurance even if full compile‑time proofs are infeasible.
      \item \textit{Trusted toolchains.}  End‑to‑end pipelines that cryptographically link a verified source graph to emitted binaries and hardware bitstreams would close the compiler gap; even partial formal checks retain value when supply chains are otherwise trusted.
      \item \emph{Hierarchical IR differencing and proof-carrying code.} Coupling static taint analysis (e.g.\ \emph{Batch Isolation Checker}’s Monoid-based proof~\cite{Kuchler2025BatchSteal}) with lowering passes can certify non-interference at compile time.
    \end{itemize}
    
    \noindent
    Bridging HPC pipelines with rigorous proofs thus remains a grand challenge, one that will require tight collaboration among ML-security specialists, compiler/toolchain engineers, and formal-methods researchers.
    
\subsection{Addressing Multi‑Path, Distributed, or Obfuscated Triggers}
\label{sec:future-multipath}

\textbf{Problem statement:} Modern backdoor designers increasingly distribute trigger logic across multiple computational paths, modalities, or time steps. CoT prompts, hidden gating sub‑layers, and cross‑modal token–image combinations can all serve as stealthy activation keys that no single neuron, branch, or dataset sample exposes \cite{Zhao2025LLMBackdoors}. Benchmark suites remain heavily skewed toward single‑branch image classifiers, leaving multimodal and multi‑path threats under‑represented.

\noindent
\textbf{Why current defenses fall short}
\begin{itemize}[leftmargin=1.5em,itemsep=2pt]
    \item \emph{Isolation bias.}  Structural detectors score each subgraph independently and thus miss backdoors that only activate when several branches co-fire in concert.  
    \item \emph{Evasive distributed triggers.}  Distributed-gating networks~\cite{Langford2025, Bober-Irizar2023} and \emph{LADDER’s} frequency-domain vision triggers~\cite{liu2025ladder} adapt to pruning and still reach $>\!90\%$ attack success while keeping clean accuracy intact.  
    \item \emph{CoT hijacks withstand consistency checks.} Lightweight tuning pipelines such as \emph{BadChain} \cite{xiang2024badchain}, \emph{DarkMind} \cite{DarkMind2025}, and \emph{ShadowCoT} \cite{Zhao2025ShadowCoT} embed triggers deep in the reasoning pathway, defeating feature-similarity detectors such as \emph{Neural Cleanse} \cite{Wang2019} and \emph{MNTD} \cite{Kolouri2021MNTD}.
    \item \emph{Modal gap.}  No public benchmark yet evaluates triggers that fuse text with image or audio modalities, leaving multi-modal gates untested~\cite{NISTTrojAI,zeng2024clibe}.
    \item \emph{Batch-context leaks.}
      A hidden path can pass information across examples inside one inference batch, enabling data theft and peer-output manipulation  ~\cite{Kuchler2025BatchSteal}.

\end{itemize}

\noindent
\newpage
\textbf{Promising research directions}
\begin{itemize}[leftmargin=1.5em,itemsep=2pt]
  \item \emph{Graph‑level constraint solving.}  Combine static dataflow graphs with dynamic activation traces to enumerate improbable multi‑branch co‑activations; graph neural networks (GNNs) could score subgraph tuples rather than single nodes.
  \item \emph{Composite fuzzing for multimodal pipelines.}  Iteratively synthesize paired inputs (e.g., prompt-image) that search the joint space for trigger conditions; coverage‑guided feedback can steer toward rare co‑activation states.
  \item \emph{Robust multi‑branch mitigation.}  Adversarial unlearning or mask‑and‑fine‑tune loops must prune all partial triggers.  Layer‑wise causal tracing, combined with sparsity‑aware retraining (§\ref{sec:mitigation-pruning}), may disable distributed gates without catastrophic accuracy loss.
  \item \emph{Evolving benchmarks.}  A living “arms‑race track” that adds the latest multi‑modal backdoor designs each release cycle would provide a moving target for defenders and avoid the staleness that plagues current benchmark corpora.
\end{itemize}

\begin{table}[htbp]
  \caption{\textbf{Benchmark coverage and key gaps}}
  \label{tab:benchmark_coverage_gap}
  \centering
  \footnotesize
  \begin{tabularx}{\linewidth}{P{3cm} Y Y}
    \toprule
    \textbf{Benchmark} & \textbf{Included} & \textbf{Missing (Critical Gaps)} \\
    \midrule
    TrojAI (IARPA) &
      Many poisoned \& weight-backdoored models; standardized detection metrics &
      No multi-branch or compiler-level attacks; primarily data- and weight-focused \\ \midrule
    BackdoorBench &
      Unified framework for comparing defenses on standard image datasets; basic single-layer structural triggers &
      No compiler-tampering scenarios; lacks complex multi-trigger architectures; primarily designed around visible triggers \\ \midrule
    NeurIPS Trojan Detection Challenges &
      Primarily data- and parameter-based triggers in competition settings &
      No explicit inclusion of structural or compiler-inserted scenarios \\ \midrule

BackdoorLLM (2024) &
  Prompt- and fine-tune-based LLM backdoors; open-source evaluation scripts &
  No structural/graph or compiler-level attacks; structural attacks: None \\ \midrule

TrojanZoo (2022) &
  Static image datasets and code for backdoor research; easy defence benchmarking &
  No dynamic updates; no compiler-time or multi-branch attacks; structural attacks: None \\

    \bottomrule
  \end{tabularx}
\end{table}

\subsection{NAS and AutoML Vulnerabilities}
\label{sec:future-automl}

\textbf{Problem statement.}
NAS and broader AutoML pipelines can embed backdoor logic if an attacker tampers with the reward function, the search code, or the compilation stage.  Because these systems often output irregular, massive graphs, hidden submodules may be indistinguishable from legitimate skip connections \cite{pang2022,pang2023dark}.

\noindent
\textbf{Why current defenses fall short}
\begin{itemize}[leftmargin=1.5em,itemsep=2pt]
    \item \emph{Opaque search traces.}  Commercial NAS services rarely provide a verifiable log of evaluated candidates, so reviewers cannot tell whether a suspicious gating op was present during architecture selection.
    \item \emph{Reward-function poisoning.}  An adversary can bias the search toward architectures that conceal malicious subgraphs while still meeting published accuracy targets.
    \item \emph{Compiler-time drift.}  Even if the final high-level graph looks clean, low-level IR (e.g., ONNX) can be patched post-training, re-introducing hidden branches before deployment.
    \item \emph{Scale barrier.}  Static scanners struggle with the tens of thousands of candidate graphs  produced in a single large-scale search, making exhaustive auditing impractical.
\end{itemize}

\noindent
\textbf{Promising research directions}
\begin{itemize}[leftmargin=1.5em,itemsep=2pt]
    \item \emph{Cryptographically verifiable search logs.} Hash-chained records of every candidate (graph, seed, reward) would let third parties confirm that the deployed model derives from an untampered search trace.
    \item \emph{IR differencing across compilation steps.} Automated checks can diff intermediate representations against the audited source graph, flagging any new ops inserted after the AutoML phase.
    \item \emph{White-list–based architecture validation.} Require search controllers to declare an explicit list of allowable modules; any unexplained layer (e.g., unconventional gating) fails certification.
    \item \emph{Search-time anomaly detection.}  Lightweight graph fingerprints computed during NAS can spot sudden structural deviations, catching backdoor candidates before expensive training finishes.
\end{itemize}
\noindent These measures should eventually align with supply-chain hardening (see §\ref{sec:mitigation-supplychain}) so that a verified AutoML output is not silently altered by later compilation or hardware integration.

\subsection{Compiler/Hardware Synergy and Post-Training Backdoor Re-Introduction}
\label{sec:future-compiler-hw}

\textbf{Problem statement.}
Even after a model passes structural scans, a malicious compiler or bitstream generator can re-insert excised subgraphs, and hardware Trojans can activate dormant gates at runtime \cite{warnecke2023evil}. Recent work such as \emph{ImpNet}~\cite{Clifford2024} and \emph{Shadow Logic}~\cite{hiddenlayer2024} modify IR by only a few hundred bytes, silently restoring the backdoor. \emph{Batch Isolation Checker} further shows that ONNX-level graph edits can inject a backdoor long after training, enabling within-batch data leakage and output manipulation~\cite{Kuchler2025BatchSteal}. Once the binary or FPGA bitstream is shipped, software-level defenses have no visibility.

\noindent
\textbf{Why current defenses fall short}
\begin{itemize}[leftmargin=1.5em,itemsep=2pt]
    \item \emph{Visibility gap.}  Audits stop at the high-level graph; they rarely inspect the lowered IR or microcode, where a single fused op can hide a trigger.
    \item \emph{Trusting-trust scenario.}  If the entire toolchain is compromised, recompiling from source reproduces the backdoor unless the compiler itself is verified or multi-compiled.
    \item \emph{Scalability of formal proofs.}  Current verifiers cannot handle billion-parameter graphs and post-compile inserts; source-level proofs are invalidated by later IR rewrites~\cite{Pham2022CAV}.
    \item \emph{Hardware opacity.}  Bitstreams for ASICs/FPGAs are proprietary; defenders cannot easily map binary regions back to functional blocks, let alone prove the absence of stealth logic~\cite{warnecke2023evil}.
\end{itemize}

\noindent
\textbf{Promising research directions}
\begin{itemize}[leftmargin=1.5em,itemsep=2pt]
    \item \emph{Reproducible and signed builds.}  Hash-chained, deterministic compilation (e.g., Bazel’s \verb!--reproducible! flag~\cite{bazel2024reproducible} and its CI caching study~\cite{zheng2024bazelci} or container capture via \emph{ReproZip}~\cite{chirigati2016reprozip}) and diverse double-compiling~\cite{Wheeler2005} lets verifiers confirm that the deployed binary matches an audited source graph.
    \item \emph{Hierarchical IR differencing and proof-carrying code.}  Each lowering pass attaches a machine-checkable proof that no new control-flow edges reach the backdoor label set; partial proofs still add value inside a trusted supply chain. 
    \item \emph{Bitstream attestation and runtime monitors.}  Split the accelerator fabric into  whitelisted regions and apply side-channel guards or runtime hash checks for unauthorized activations~\cite{Sengupta2025HardwareTrojanML}.
    \item \emph{Cross-layer certification.}  Unify software and hardware verification by expressing both the neural graph and the accelerator netlist in a single SMT-based meta-model, enabling end-to-end equivalence checks.
\end{itemize}
\noindent
Bridging compiler transformations, hardware design, and ML graphs therefore demands a holistic, supply-chain view; defensively chaining even partial formal checks can raise the bar until tool-support scales to full proofs.

% ----------------------------------------------------------------
\subsection{Adaptive Benchmarks and Continuous Updating}
\label{sec:future-benchmarks}
% ----------------------------------------------------------------

\textbf{Problem statement.}
Most public backdoor testbeds such as \emph{TrojAI}, \emph{BackdoorBench}, \emph{TrojanZoo}, and the language-model-focused \emph{BackdoorLLM} suite \cite{wu2022backdoorbench, pang2023trojanzoo,Li2024BackdoorLLM}, remain largely static once released.  Consequently, defenses tuned to a single round of pixel or weight triggers can appear robust while collapsing against hidden-graph or compiler-time attacks that the benchmark never covered.

\noindent
\textbf{Why current defenses fall short}
\begin{itemize}[leftmargin=1.5em,itemsep=2pt]
    \item \emph{Stale attack sets.}  Few benchmarks introduce multi-modal or compiler-injected triggers, leaving entire threat classes unmeasured.
    \item \emph{Real-world drift.}  The Guardian scanner recently flagged 57 suspect ONNX graphs on Hugging Face~\cite{protectai2024PAIT}; none of those samples resemble existing challenge rounds, exposing a realism gap.
    \item \emph{HPC compilation loops.}  Production pipelines re-optimize models on every deployment.  A defense that succeeds on a frozen weight file may fail after even one hardware-aware recompilation.
\end{itemize}

\noindent
\textbf{Promising research directions}
\begin{itemize}[leftmargin=1.5em,itemsep=2pt]
    \item \emph{Annual “arms-race” tracks.}  Each round injects \emph{new} infiltration vectors (stealth gating, text–image triggers, post-training IR rewrites), forcing defenses to generalize beyond last year’s tactics.
    \item \emph{HPC-pipeline simulation.}  Benchmarks should recompile each submitted model through multiple optimization passes, rewarding detectors that remain stable across binary drift.
    \item \emph{Rolling real-world corpus tracks.}  Publicly harvested incident sets (e.g.\ the PAIT-ONNX and PAIT-TF reports from Protect AI \cite{protectai2024PAIT} plus the 269 leaking models found by Küchler\,\textit{et al.}’s scan of 1,680 ONNX graphs~\cite{Kuchler2025BatchSteal}) can supply continually refreshed challenge seeds that mirror live supply-chain threats.
    \item \emph{Live scoreboards and gap analytics.}  A web dashboard summarizing which defenses block which vectors will highlight lingering blind spots and channel research effort.
    \item \emph{Multi-modal extensions.}  Future rounds must include text–image or audio–text gates that no current benchmark yet exercises, as highlighted by Zhao \emph{et al.}’s LLM survey \cite{Zhao2025LLMBackdoors}.
    \item \emph{LLM-orchestrated defender baselines.}  Recent work shows that large language models can coordinate autonomous cyber-defense actions~\cite{Castro2025LLMDefenders}; incorporating such agents as detectors/repairers would test whether benchmarks keep pace with AI-driven defense.

\end{itemize}
\noindent
A continuously evolving benchmark ecosystem, anchored by annual arms-race tracks, realistic corpus feeds, and HPC-aware recompilation loops, offers the best hope of preventing defenses from over-fitting to yesterday’s attacks.

\subsection{Specialized Architectures: Spiking Neural Networks and Visual State-Space Models}
\label{sec:specialized-architectures}

\textbf{Problem statement.}
Architectural backdoors are no longer confined to convolutional or Transformer families. Spiking Neural Networks (SNNs) and Visual State-Space Models (VSSMs) have both been shown vulnerable to stealthy, architecture-level attacks: Sneaky Spikes embeds persistent triggers in neuromorphic timing channels \cite{Abad2024}, while BadScan tampers with  VSSM state-update blocks \cite{Deshmukh2024}. As these models proliferate in robotics, real-time control, and neuromorphic hardware, ignoring their unique threat surface leaves an ever-growing security gap.

\noindent
\textbf{Why current defenses fall short}
\begin{itemize}[leftmargin=1.5em,itemsep=2pt]
    \item \emph{Continuous-activation bias.}  Most detectors assume real-valued activations; they miss micro-timing or spike-rate perturbations that encode SNN backdoors.
    \item \emph{Hidden recurrent pathways.}  VSSMs update an internal state through learned linear systems, so a malicious matrix can stay dormant until a specific input trajectory occurs, evading static graph scans.
    \item \emph{Lack of benchmark.}  No public challenge currently includes spiking or state-space triggers; defenders have no ground truth for evaluation.
    \item \emph{Tool-chain gap.}  Neuromorphic compilers and on-device learning loops lack the signing and reproducibility features now standard in mainstream ML toolchains (§\ref{sec:future-compiler-hw}).
\end{itemize}

\noindent
\textbf{Promising research directions}
\begin{itemize}[leftmargin=1.5em,itemsep=2pt]
    \item \emph{Event-level tracing and invariants.}  Define spike-timing windows or state-flow invariants and flag inputs that violate them.
    \item \emph{Domain-specific benchmarks.}  Extend \emph{TrojAI} \cite{NISTTrojAI} or \emph{BackdoorBench} \cite{wu2022backdoorbench} with SNN timing triggers and VSSM state-matrix inserts, using an “arms-race” update cadence (§\ref{sec:future-benchmarks}).
    \item \emph{Formal abstractions for spikes and states.}  Adapt mixed-signal verification or control-theoretic reachability analysis to reason about timing-encoded or state-encoded backdoors.
    \item \emph{Cross-disciplinary}  Combine neuromorphic-hardware provenance checks with the reproducible-build approach already proposed for standard accelerators (§\ref{sec:future-compiler-hw}).
\end{itemize}

\subsection{Supply-Chain Governance, Policy, and Multi-Sector Collaboration}
\label{sec:future-policy}

\textbf{Problem statement:}
Technical countermeasures are necessary but insufficient: architectural backdoors threaten regulated sectors, healthcare, finance, autonomous vehicles, where liability, compliance, and public safety considerations dominate \cite{WEF2025Backdoor}. Without verifiable provenance and enforceable policy, a single compromised compiler stage can undermine every software-level defense.

\noindent
\textbf{Why current defenses fall short}
\begin{itemize}[leftmargin=1.5em,itemsep=2pt]
    \item \emph{Fragmented standards.}  Cryptographic signing, reproducible builds, and audit logging exist, but no cross-industry baseline mandates them for neural-network toolchains.
    \item \emph{Unclear liability.}  When hidden logic is discovered, it is ambiguous whether fault lies with the model provider, compiler vendor, or third-party library maintainer.
    \item \emph{Regulatory lag.}  Emerging laws (e.g., EU AI Act 2024 \cite{EUAIAct2024}) require risk documentation but stop short of specifying architecture-integrity attestation.
    \item \emph{Domain-specific blind spots.}  Safety-critical systems rarely perform “white-box” architectural audits, focusing instead on data or post-hoc performance metrics.
\end{itemize}

\noindent
\textbf{Promising research and policy directions}
\begin{itemize}[leftmargin=1.5em,itemsep=2pt]
    \item \emph{Signed, reproducible toolchains.}  Mandate hash-chained logs and deterministic builds for each compile step, aligning with NIST AI RMF 1.0 guidance and ISO/IEC 42001 proposals \cite{NIST2023AI_RMF}.
    \item \emph{Sector-specific integrity audits.}  Regulators could require  provable “white-box” scans of autonomous-vehicle or medical models before type approval, similar to functional-safety testing today.
    \item \emph{Shared liability frameworks.}  Clarify contractual obligations so that model providers, tool-vendors, and cloud operators share responsibility for preventing compiler-level backdoors.
    \item \emph{Third-party certification.}  Create accredited services that certify neural architectures against malicious subgraph insertion, mirroring SOC 2 or Common Criteria for traditional software.
    \item \emph{Cross-sector working groups.}  Foster collaboration among neuromorphic, HPC, and policy communities to translate supply-chain lessons from conventional software into the ML domain.
\end{itemize}

\noindent
Embedding these governance layers into the ML lifecycle ensures that technical defenses such as the reproducible-build pipeline in §\ref{sec:future-compiler-hw}, are backed by enforceable policy, shifting architectural-backdoor mitigation from best-effort practice to industry norm.

\subsection{Proposed Research Roadmap}
\label{sec:future-roadmap}

\begin{table}[htbp]
  \caption{\textbf{Short-Term vs.\ Long-Term Research Goals for Architectural Backdoor Security}}
  \label{tab:future_roadmap}
  \centering
  \footnotesize                        % ACM-approved small size
  % two fixed-width ragged columns and one flexible ragged column
  \begin{tabularx}{\columnwidth}{P{3.2cm} P{4.0cm} Y}
    \toprule
    \textbf{Challenge} & \textbf{Short-Term Goals} & \textbf{Long-Term Goals}\\
    \midrule
    \textbf{Large-Scale Verification} &
      Develop HPC-friendly partial proofs; compositional checks on sub-networks &
      Full cross\-layer verification that scales to trillion-parameter LLMs or HPC pipelines \\
    \midrule
    \textbf{Multi-Path, Distributed Triggers} &
      Integrate multi-branch infiltration into existing detection tools; refine gating instrumentation &
      Deploy constraint-based or real-time gating checks in industrial systems; unify with formal proofs of activation patterns \\
    \midrule
    \textbf{Malicious NAS/AutoML} &
      Cryptographically log each architecture search step; IR-level differencing &
      Fully transparent AutoML pipelines with forced architecture declarations and standardized subgraph scanning \\
    \midrule
    \textbf{Compiler/Hardware Backdoor Re-Introduction} &
      Expand reproducible builds; cryptographic signing at compile time &
      End-to-end synergy across compiler, hardware bitstreams, and ML frameworks; hardware-level concurrency checks \\
    \midrule
    \textbf{Adaptive Benchmarks} &
      Add pilot structural infiltration to TrojAI \cite{NISTTrojAI} or BackdoorBench \cite{wu2022backdoorbench} &
      Establish continuous, arms-race style expansions for HPC or multi-modal infiltration \\
    \midrule
    \textbf{Supply-Chain Governance and Policy} &
      Introduce code signing and domain-specific compliance checks &
      Formal policy frameworks ensuring secure pipelines from architecture code to final hardware implementations \\
    \bottomrule
  \end{tabularx}
\end{table}

\noindent
Table~\ref{tab:future_roadmap} explicitly outlines actionable short-term goals and ambitious long-term objectives to guide researchers, practitioners, and policymakers. For instance, benchmark maintainers should introduce at least one multi-branch or compiler-inserted Trojan scenario in the next round of \emph{TrojAI} or \emph{BackdoorBench}. For longer-term grand challenges, such as full cross-layer verification of trillion-parameter models, new theoretical frameworks may be required that constrain model architecture to enable tractable verification without performance degradation. Such challenges necessitate broad, interdisciplinary collaboration. By coordinating short-term efforts (e.g., cryptographic logs, partial compositional proofs, pilot structural infiltration in existing benchmarks) with these long-term objectives (full HPC-scale verification, continuous pipeline scanning, multi-modal backdoor expansions), the field can gradually construct a robust, end-to-end security strategy.

\subsection{Summary}
\label{sec:future-conclude}

Architectural backdoors, ranging from multi-path triggers to NAS-biased topologies, expose gaps in parameter-centric defences; priorities now are (i) scalable formal proofs for multi-branch HPC models, (ii) transparent and logged AutoML/compilation pipelines, (iii) hardened supply chains with reproducible builds, (iv) adaptive benchmarks that evolve with new infiltration patterns, and (v) extending tests to LLM and multi-modal systems. Ultimately, progress in these open challenges can raise the bar for architectural backdoor security across software and hardware boundaries. The community-encompassing ML security researchers, compiler engineers, hardware designers, and policy/regulatory experts-must collaborate to build end-to-end, verified pipelines that deter architectural backdoors even in large-scale, rapidly evolving AI ecosystems. By integrating supply-chain assurance, HPC-scale verification, and iterative benchmark expansions, we can move closer to a future where malicious subgraphs, multi-branch gating, and post-training backdoor reintroductions are consistently detected and neutralized in practice.

\section{Conclusion and Outlook}
\label{sec:conclusion}

\begin{figure*}[t]
\centering
\resizebox{\textwidth}{!}{%
\begin{tikzpicture}[
  detect/.style  ={draw, rounded corners, fill=teal!25,
                   minimum width=3cm, align=center, font=\footnotesize},
  mit/.style      ={draw, rounded corners, fill=orange!30,
                   minimum width=3cm, align=center, font=\footnotesize},
  bench/.style    ={draw, rounded corners, fill=green!30,
                   minimum width=3cm, align=center, font=\footnotesize},
  future/.style   ={draw, rounded corners, fill=violet!30,
                   minimum width=3cm, align=center, font=\footnotesize},
  sub/.style      ={draw, rounded corners, fill=gray!15,
                   minimum width=2.2cm, align=center, font=\footnotesize},
  core/.style     ={draw, rounded corners, fill=gray!30,
                   minimum width=3.2cm, align=center, font=\footnotesize},
  arrow/.style    ={->, >=Stealth, thick},
  node distance   =0.8cm and 1.6cm,
  on background layer,
  band/.style     ={fill opacity=.12, draw=none, rounded corners}
]
\usetikzlibrary{fit,backgrounds}

% --- central node & main themes ---------------------------------
\node[core] (core) {Architectural\\Backdoor Road-map};

\node[detect, above=of core] (detect) {Detection};
\node[mit,    below=of core] (mit)    {Mitigation};
\node[bench,  left =of core] (bench)  {Benchmarks};
\node[future, right=of core] (future) {Open Problems};

% --- detection sub-nodes ----------------------------------------
\node[sub, above left =of detect]  (static) {Static Diff};
\node[sub, above       =of detect] (dyn)    {Dyn.\ Probing};
\node[sub, above right =of detect] (form)   {Formal Verif.};

% --- mitigation sub-nodes ---------------------------------------
\node[sub, below left =of mit]  (prune) {Subgraph\\Excision};
\node[sub, below       =of mit] (unl)   {Unlearning};
\node[sub, below right =of mit] (supp)  {Supply-chain};

% --- benchmark sub-nodes ----------------------------------------
\node[sub, above left =of bench] (legacy){Extend\\TrojAI/BB};
\node[sub, left        =of bench] (repair){Repair\\Metrics};
\node[sub, below left =of bench] (arms)  {Arms-race\\Tracks};

% --- future sub-nodes -------------------------------------------
\node[sub, above right =of future] (scale){LLM-scale\\Verif.};
\node[sub, right        =of future] (multi){Multi-Trigger};
\node[sub, below right =of future] (hw)   {HW Trojans};

% --- pastel background bands (fit around each theme cluster) ----
%\node[band, fill=teal!40,  fit=(detect) (static) (dyn) (form)] {};
%\node[band, fill=orange!40,fit=(mit)    (prune) (unl) (supp)] {};
%\node[band, fill=green!40, fit=(bench)  (legacy) (repair) (arms)] {};
%\node[band, fill=violet!40,fit=(future) (scale) (multi) (hw)] {};

\node[band, fill=teal!40,  fit=(static) (dyn) (form)] {};
\node[band, fill=orange!40,fit=(prune) (unl) (supp)] {};
\node[band, fill=green!40, fit=(legacy) (repair) (arms)] {};
\node[band, fill=violet!40,fit=(scale) (multi) (hw)] {};

% --- arrows -----------------------------------------------------
\foreach \m in {detect,mit,bench,future}
    \draw[arrow] (core) -- (\m);
\foreach \s in {static,dyn,form}
    \draw[arrow] (detect) -- (\s);
\foreach \s in {prune,unl,supp}
    \draw[arrow] (mit) -- (\s);
\foreach \s in {legacy,repair,arms}
    \draw[arrow] (bench) -- (\s);
\foreach \s in {scale,multi,hw}
    \draw[arrow] (future) -- (\s);
\end{tikzpicture}}
\caption{\textbf{Research roadmap for architectural-backdoor security.}The figure groups open work into four color-banded themes: \textcolor{teal!70!black}{teal} = \emph{detection}, \textcolor{orange!80!black}{orange} = \emph{mitigation}, \textcolor{green!70!black}{green} = \emph{benchmarking}, and \textcolor{violet!70!black}{violet} = \emph{open problems} \cite{DarkMind2025,Sengupta2025HardwareTrojanML,EUAIAct2024}.}

\label{fig:conclusion_map}
\Description{A central gray box labelled “Architectural Backdoor Road-map” sits in the middle of the page.  Four colored bands radiate outwards: a teal horizontal band above the center (Detection), an orange horizontal band below (Mitigation), a vertical green band on the left (Benchmarks), and a vertical violet band on the right (Open Problems). Each band contains its theme node plus three smaller sub-task boxes positioned around it; thin black arrows run from the theme node to every sub-task.  
Additional black arrows connect the four colored theme nodes back to the central roadmap box.}

\end{figure*}
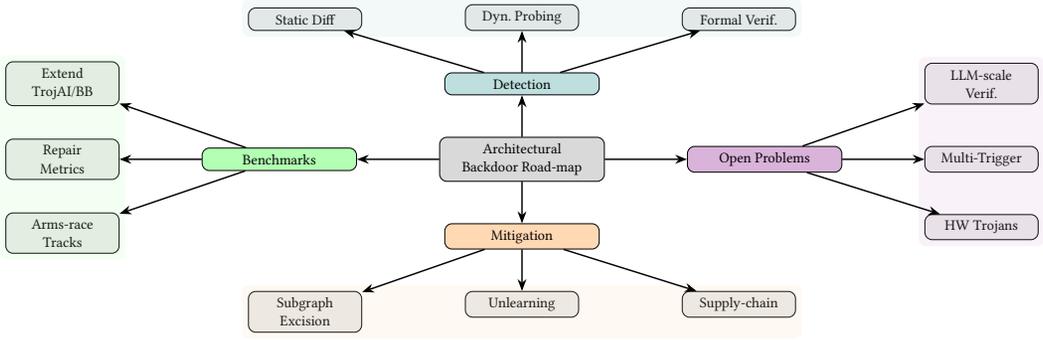

Figure~\ref{fig:conclusion_map} groups the remaining research challenges into four color-coded themes: detection, mitigation, benchmarking, and open problems which correspond to successive stages in the machine-learning supply chain. The discussion below walks clockwise through those bands.

\noindent
\textbf{\textcolor{teal!70!black}{Detection (teal band):}}
Static differencing, dynamic probing, and formal verification form the first checkpoint. Yet multi-branch triggers such as Langford’s distributed gates\cite{Langford2025}, \emph{DarkMind’s} latent CoT hijacks \cite{DarkMind2025}, and \emph{BadChain’s} lightweight tuning \cite{xiang2024badchain} routinely evade single-path heuristics.  SMT-based provers stall on billion-parameter graphs \cite{Pham2022CAV}, and compiler-time insertions like \emph{Shadow Logic} \cite{hiddenlayer2024} compromise even “clean” source graphs. Scalable, graph-wide reasoning therefore remains open. 

\noindent
\textbf{\textcolor{orange!80!black}{Mitigation (orange band):}} Sub-graph excision, unlearning, or signed tool-chains can lower attack success, but a poisoned build system can silently re-insert the backdoor (\emph{ImpNet}~\cite{Clifford2024}) or activate a dormant hardware Trojan \cite{Sengupta2025HardwareTrojanML}.  Supply-chain hardening, hash-chained, reproducible builds, container capture, and diverse double-compiling \cite{Wheeler2005}, bind verified graphs to emitted binaries and close the “trusting-trust” loop.

\noindent
\textbf{{\textcolor{green!70!black}{Benchmarking (green band):}}}
Legacy challenges \emph{(TrojAI, BackdoorBench, TrojanZoo)} seldom test structural attacks, and no round yet stresses cross-modal gates highlighted by Zhao \emph{et al.} \cite{Zhao2025LLMBackdoors}.  Hub scanners such as \emph{CLIBE} model \cite{zeng2024clibe} and the PAIT corpus \cite{protectai2024PAIT} reveal stealthy exploits already in the wild.  A rolling, arms-race track, augmented with explicit repair metrics to grade mitigation quality, would turn benchmarks into a living barometer rather than a static checklist.

\noindent
\textbf{\textcolor{violet!70!black}{Open problems (violet band):}} LLM-scale verification, multi-trigger reasoning, and hardware–compiler synergy are unresolved frontiers.  The EU AI Act \cite{EUAIAct2024} and NIST AI RMF 1.0 \cite{NIST2023AI_RMF} require provenance but still lack architecture-integrity clauses. Consensus across research, industry, and policy is needed before HPC and multi-modal deployments become the default.

\subsection{Call to action}
\begin{itemize}[leftmargin=1.5em,itemsep=2pt]
  \item \emph{Detect}: build graph-level solvers that scale to trillion-parameter LLMs and expose multi-branch activations.
  \item \emph{Mitigate}: embed reproducible builds, IR differencing, and runtime attestation in CI/CD pipelines.
  \item \emph{Benchmark}: launch yearly arms-race rounds, scored on detection and repair.
  \item \emph{Govern}: codify architectural-integrity attestation and shared liability in emerging AI standards and policy.
\end{itemize}

\noindent
Architectural backdoor defense, therefore, spans design, compilation, and deployment, and succeeds only through joint effort of data scientists, software engineers, hardware designers, security experts, and regulators acting in concerted collaboration. As HPC-scale deep learning accelerates, closing these structural loopholes now is pivotal to safeguarding future AI systems.  We hope this survey catalyzes that multi-disciplinary collaboration.

\bibliographystyle{ACM-Reference-Format}
\bibliography{references}

\end{document}